\documentclass[a4paper,11pt]{article}
\pdfoutput=1
\usepackage{jcappub}
\usepackage[utf8]{inputenc}
\usepackage[english]{babel}
\usepackage{physics}
\usepackage{bbold}
\usepackage{amssymb,amsmath,amsfonts}
\usepackage{csquotes}
\usepackage{fancyvrb}
\usepackage{tensor}
\usepackage{layout}
\usepackage[normalem]{ulem}
\usepackage{booktabs}

\usepackage{multirow}
\usepackage{verbatim}
\usepackage{comment}
\usepackage{scalefnt}
\usepackage[usenames,dvipsnames]{xcolor}

\usepackage{graphicx,color}

\usepackage{orcidlink}

\usepackage{subcaption}
\usepackage{soul,float}

\usepackage{lipsum}
\usepackage{cancel}

\usepackage{listings}
\usepackage{xcolor}

\definecolor{codegreen}{rgb}{0,0.6,0}
\definecolor{codegray}{rgb}{0.5,0.5,0.5}
\definecolor{codepurple}{rgb}{0.58,0,0.82}
\definecolor{backcolour}{rgb}{0.95,0.95,0.92}

\lstdefinestyle{mystyle}{
    backgroundcolor=\color{backcolour},   
    commentstyle=\color{codegreen},
    keywordstyle=\color{magenta},
    numberstyle=\tiny\color{codegray},
    stringstyle=\color{codepurple},
    basicstyle=\ttfamily\footnotesize,
    breakatwhitespace=false,         
    breaklines=true,                 
    captionpos=b,                    
    keepspaces=true,                 
    numbers=left,                    
    numbersep=5pt,                  
    showspaces=false,                
    showstringspaces=false,
    showtabs=false,                  
    tabsize=2,
    inputencoding = utf8,  
    extendedchars = true,  
    literate      =        
      {á}{{\'a}}1  {é}{{\'e}}1  {í}{{\'i}}1 {ó}{{\'o}}1  {ú}{{\'u}}1
      {Á}{{\'A}}1  {É}{{\'E}}1  {Í}{{\'I}}1 {Ó}{{\'O}}1  {Ú}{{\'U}}1
      {à}{{\`a}}1  {è}{{\`e}}1  {ì}{{\`i}}1 {ò}{{\`o}}1  {ù}{{\`u}}1
      {À}{{\`A}}1  {È}{{\'E}}1  {Ì}{{\`I}}1 {Ò}{{\`O}}1  {Ù}{{\`U}}1
      {ä}{{\"a}}1  {ë}{{\"e}}1  {ï}{{\"i}}1 {ö}{{\"o}}1  {ü}{{\"u}}1
      {Ä}{{\"A}}1  {Ë}{{\"E}}1  {Ï}{{\"I}}1 {Ö}{{\"O}}1  {Ü}{{\"U}}1
      {â}{{\^a}}1  {ê}{{\^e}}1  {î}{{\^i}}1 {ô}{{\^o}}1  {û}{{\^u}}1
      {Â}{{\^A}}1  {Ê}{{\^E}}1  {Î}{{\^I}}1 {Ô}{{\^O}}1  {Û}{{\^U}}1
      {œ}{{\oe}}1  {Œ}{{\OE}}1  {æ}{{\ae}}1 {Æ}{{\AE}}1  {ß}{{\ss}}1
      {ç}{{\c c}}1 {Ç}{{\c C}}1 {ø}{{\o}}1  {Ø}{{\O}}1   {å}{{\r a}}1
      {Å}{{\r A}}1 {ã}{{\~a}}1  {õ}{{\~o}}1 {Ã}{{\~A}}1  {Õ}{{\~O}}1
      {ñ}{{\~n}}1  {Ñ}{{\~N}}1  {¿}{{?`}}1  {¡}{{!`}}1
      {°}{{\textdegree}}1 {º}{{\textordmasculine}}1 {ª}{{\textordfeminine}}1
}

\usepackage{tcolorbox}
\newtcolorbox{mybox}[1]{
		colback=black!5!white, 
		colframe=black, 
		fonttitle=\bfseries,title=#1}

\usepackage{hyperref}

\newcommand{\be}{\begin{eqnarray}}
\newcommand{\ee}{\end{eqnarray}}
\newcommand{\ba}{\begin{array}}
\newcommand{\ea}{\end{array}}

\title{On the use of the Derivative Approximation for Likelihoods for Gravitational Wave Inference}

\author[a]{Josiel Mendonça Soares de Souza\orcidlink{0000-0003-1552-0095},}
\author[a, b, c, d]{Miguel Quartin\orcidlink{0000-0001-5853-6164}}

\affiliation[a]{Instituto de Física, Universidade Federal do Rio de Janeiro, 21941-972, Rio de Janeiro, RJ, Brazil}

\affiliation[b]{Centro Brasileiro de Pesquisas Físicas, 22290-180, Rio de Janeiro, RJ, Brazil}

\affiliation[c]{Observatório do Valongo, Universidade Federal do Rio de Janeiro, 20080-090, Rio de Janeiro, RJ, Brazil}

\affiliation[d]{PPGCosmo, Universidade Federal do Espírito Santo, 29075-910, Vitória, ES, Brazil}

\abstract{
Posterior inference on the more than a dozen parameters governing a gravitational wave (GW) event is challenging. A typical MCMC analysis can take around $100$ CPU hours, and next generation GW observatories will detect many thousands of events. Here we present a thorough comparison of the accuracy and computational cost of the Fisher Matrix, Derivative Approximation for Likelihoods (DALI) and traditional MCMC methods. We find  that using DALI, which extends the traditional Fisher Matrix (FM) method to higher orders, allows for a good approximation of the posterior with a $55$ times smaller computational cost, and that the cost-benefit of the doublet-DALI is better than that of the triplet-DALI. We also show that the singlet-DALI, a hybrid MCMC-Fisher method, is much more accurate than the traditional FM and 10 times faster than the doublet-DALI. A large effort has been invested in forecasting the science case of different detector configurations, and the ability of making fast yet accurate estimations of the posteriors is an important step forward. We also introduce version \texttt{1.0} of the public \texttt{GWDALI} code, which incorporates automatic differentiation, modern waveforms and an optimized parameter decomposition.
}

\begin{document}

\maketitle

\section{Introduction}

A decade has passed since the direct detection of the first gravitational wave (GW). The field has since made fast progress on many different fronts. In terms of detections, the LIGO-Virgo-Kagra collaboration has recently released the fourth GW catalog~\cite{LIGOScientific:2025slb}, which contains 138 binary black hole (BBH), 3 neutron star-black hole (NSBH), and 2 binary neutron star (BNS) coalescences when using a conservative threshold of false alarm rate (FAR) smaller than 0.25 per year~\cite{LIGOScientific:2025pvj}.

The main proposals for the next generation of GW observatories are the Einstein Telescope~\cite{Punturo:2010zz} and the Cosmic Explorer~\cite{Reitze:2019iox}.
Both should have much improved sensitivity and probe lower inspiral frequencies. The Einstein Telescope alone should increase the number of binary coalescence detections by at least two orders of magnitude and in a much larger range of redshifts and masses~\cite{Iacovelli:2022bbs}, vastly increasing the expected science output~\cite{Abac:2025saz}.

Each GW event is described by several parameters. In the simplest case, for BBH in circular orbits and where both spins are assumed to be aligned, the number of parameters needed to describe the waveform is eleven. An accurate calculation of the posterior of these parameters can be achieved using Markov-Chain Monte Carlo (MCMC) as well as nested sampling based techniques. These are, however, costly, in particular due to the non-Gaussian nature of the posterior. Added to this difficulty is the need to repeat this analysis for the tens of thousands of expected events. It is therefore interesting to explore possible shortcuts in the calculation of these posteriors.

In recent years, the community has put forth great effort to build Fisher Matrix (FM) codes for GW forecasts~\cite{Borhanian:2020ypi,Dupletsa:2022scg,Iacovelli:2022mbg,Begnoni:2025oyd}. Fisher matrices are often used in science for this purpose, since it allows very fast estimation of confidence levels and marginalization, at the cost of assuming that the posteriors are Gaussian in the parameters. For GW parameter estimation, in particular, it has been in use for over 30 years~\cite{Cutler:1994ys}. However, when considering co-located detectors or two spatially separated detectors, the GW posteriors are not approximately Gaussian, even for high-SNR sources, and the corresponding FM are often nearly singular. As such, its inverse, which is the covariance matrix, is plagued by numerical instabilities~\cite{Vallisneri:2007ev,Rodriguez:2013mla}. This motivates the community to look for alternative methods. One possibility is to try to reparametrize the waveforms using parameters that exhibit stronger Gaussianity and, in particular, reduced multimodality~\cite{Roulet:2022kot}. Another is to separate the parameter space into two sub-groups, and treat one of these analytically~\cite{Chassande-Mottin:2019nnz,Mancarella:2024qle}.  Effective Fisher matrices were also proposed as an alternative~\cite{Cho:2012ed}. Finally, one can also use the FM to build an exact Gaussian likelihood which can then be sampled with Monte Carlo or similar methods~\cite{desouza:2023ozp,Dupletsa:2024gfl}. 

Here, instead, we investigate the performance of the DALI (Derivative Approximation for Likelihoods) method proposed in~\cite{Sellentin:2014zta} and further developed in~\cite{Sellentin:2015axa}. It consists of a higher-order expansion of the likelihood around the best fit. A naïve Taylor expansion breaks down beyond the second order used in the Fisher Matrix methodology, as the posteriors lose their positive-definiteness and normalizability. It was shown in~\cite{Sellentin:2014zta} that, if one groups the series expansion terms by the order of the derivative terms instead as by the powers of the perturbation around the best fit, both properties of a well-behaved distribution are recovered. 

In the GW-DALI approach, if the $N$ parameters are described by $\boldsymbol{\theta} = \{\theta_1, \dots, \theta_N\}$, we write the log likelihood, as an expansion around the best fit (BF) as follows:
\begin{align}\label{Eq:DALI}
    \log\mathcal{L}  \;=\;
    & \log\mathcal{L}_{\rm BF} - \left[\frac{1}{2}\sum_{i,j}\langle \partial_i h| \partial_j h \rangle \Delta \theta^{ij}\right]_{\rm BF} \nonumber\\
    &  { - \left[\frac{1}{2}\sum_{i,j,k}\langle \partial_i h | \partial_j\partial_k h \rangle\Delta\theta^{ijk}
         +\frac{1}{8}\sum_{i,j,k,l}\langle \partial_i\partial_j h | \partial_k\partial_l h \rangle\Delta\theta^{ijkl}\right]_{\rm BF}} \nonumber \\
    & {- \left[\frac{1}{6}\sum_{i,j,k,l}\langle \partial_i h | \partial_j\partial_k\partial_l h \rangle\Delta\theta^{ijkl}
         +\frac{1}{12}\sum_{i,...,m}\langle \partial_i\partial_j h | \partial_k\partial_l\partial_m h \rangle\Delta\theta^{ijklm}\right.}\nonumber \\
    &\quad\;\; \left. + \,\frac{1}{72} \sum_{i,...,n}\langle \partial_i\partial_j\partial_k h | \partial_l\partial_m\partial_n h \rangle\Delta\theta^{ijklmn} \right]_{\rm BF}\,,
\end{align}
where $\Delta\theta^i\equiv \theta^i-\theta^i_0$ and $\Delta\theta^{i\dots k} \equiv \Delta\theta^i \dots \Delta\theta^k$ \cite{Wang:2022kia}. The first term in brackets is the standard FM. The second, is called the doublet term, and contains all second-order derivatives terms but no higher-order ones. Finally, the last term in brackets is called the triplet, and contains the third-order derivatives terms, but again no higher ones. The expansion could be continued if desired. As was shown in~\cite{Sellentin:2014zta}, stopping the series at the doublet level allows the reconstruction of mildly non-Gaussian posteriors, whereas stronger non-Gaussianities require including the triplet term for a better approximation. Clearly, for highly non-Gaussian posteriors the expansion is expected to break down. Moreover, the number of independent terms increase substantially at each order: $N(N+1)/2$ for the doublet and $(N^3+3N^2+2N)/6$ for the triplet. One must thus weight the computational cost-benefit to decide if it is worth or not to add the triplet. In any case, once the derivatives are computed, the likelihood becomes very fast to evaluate, and a complete exploration of the parameter space with, say, an MCMC method, proceeds much faster.

We make a quick remark to draw attention that the DALI method could also be used directly on the posteriors, instead of the likelihoods. In practice, however, there is little benefit of doing so, as the priors adopted are often analytical and simple to implement on top of the DALI approximation. Multiplying the priors directly to the DALI also allows using priors with sharp boundaries, such as the often used top-hat priors.

In the context of GW, the DALI method was first investigated in~\cite{Wang:2022kia}, and promising results were found in the reanalysis of real data when using either the doublet or the triplet. Further research on the topic was conducted by~\cite{desouza:2023ozp}, where a preliminary version of the public \texttt{GWDALI} code was developed and tested with the simple \texttt{TaylorF2} waveform \cite{Damour:2000zb}. Here, we developed a substantially enhanced version of \texttt{GWDALI}, which is now being released as version 1.0. The new version includes the ability to use auto-differentiation techniques~\cite{autodiff1964} using the \texttt{JAX} library~\cite{jax2018github}, it allows reparametrizations, and adds more modern waveforms (in particular \texttt{IMRPhenomHM}~\cite{phenomHM} and its predecessors), which were all rewritten from the code \texttt{LAL} (LIGO-Virgo-Kagra Algorithm Library)~\cite{lalsuite} to pure python via \texttt{JAX}. Auto-differentiation, or simply auto-diff, circumvents the fact that numerical derivatives are intrinsically noisy, and subject to numerical errors which are hard to control, and does away with the need to choose and calibrate the finite step size. This is especially important for higher derivatives, which the DALI method relies on. The use of auto-diff has been incorporated in previously developed FM codes~\cite{Iacovelli:2022mbg,Begnoni:2025oyd}. Nevertheless, we stress that while auto-diff removes a potential point of failure for numerical derivatives, it does not solve the unreliability of FM in GW analyses, and we find many instances where the FM predictions are substantially different from the exact likelihood.

In this paper we use the new \texttt{GWDALI} release to probe in detail the suitability of the DALI approach in GW. We use the \texttt{IMRPhenomHM} approximant as our fiducial waveform and assume aligned spins. This reduces the parameter-space to eleven dimensions, and at the same time avoids the computation of more complex waveforms such as, e.g., \texttt{IMRPhenomXPHM}~\cite{Pratten:2020ceb} or \texttt{SEOBNRv5PHM}~\cite{Ramos-Buades:2023ehm}, which include spin precession. We explore how well the doublet-DALI and triplet-DALI approximate the posteriors for 300 BBH injections with signal-to-noise (SNR) ratio larger than 8 for the Einstein Telescope (see Section~\ref{sec:ET}), comparing the results with full exact\footnote{Thoughout this text we will dub Exact a posterior which was obtained through a full Bayesian parameter estimation using a converged MCMC sample.} MCMC posteriors for the same 300 events, as well as with the FM. We also compare the computational cost of using DALI at the doublet or triplet level when compared to traditional MCMC methods.

\subsection{Kullback-Leibler and Jensen-Shannon divergences}

When comparing approximate methods for likelihoods or posteriors, one can use different quantities to quantify the quality of the approximation. If one simply wants to forecast the distribution of marginalized uncertainties in each parameter, one can simply compare 1D histograms of the standard deviations in each parameter. If, instead, one is interested on how unbiased the final contours are, when compared with the exact ones, two other quantities are of interest: the Kullback-Leibler Divergence (KLD, also called relative entropy)~\cite{kullback1951information} and the Jensen-Shannon Divergence (JSD)~\cite{Lin-JSD-1991}. 

The KLD is an asymmetric divergence from a reference distribution $P$ to another $Q$, and measures how much extra information is needed to go from $P$ to $Q$. It is defined as
\begin{equation}
    D_{KL}(P|Q) = \int P(\boldsymbol{\theta}) \log\left[\frac{P(\boldsymbol{\theta})}{Q(\boldsymbol{\theta})}\right] \dd^n\boldsymbol{\theta} \,.
\end{equation}
The KLD is zero only when both distributions coincide completely. Otherwise it takes on positive values, which become larger as the discrepancy increases. It is the best suited quantity when predictive quality is the main concern. For instance, for real GW data, in order to inform follow-up telescopes about the GW localization with an approximate posterior, the approximation with the lowest KLD (computed from the exact to the approximation) would be the best one to use. Besides the case of approximate posteriors, the KLD has also been used as a basis for comparing the agreement between different cosmological datasets~\cite{Grandis:2015qaa,Seehars_2016,Mello:2024tor}.

On the other hand, for a simple agreement score between two distributions, the JSD is more appropriate. It is defined as
\begin{equation}
    D_{JS} = \frac{1}{2}\Big[ D_{KL}(P|M) + D_{KL}(Q|M) \Big],
\end{equation}
where $M \equiv (P+Q)/2$ is the mixture distribution of $P$ and $Q$. It is thus based on the KLD, but it is symmetric, and smoother. It is therefore numerically more stable and less susceptible to disagreements on the tails of the distributions. Here, since we are dealing with simulated data and mainly interested in getting reliable forecasts, we will quantify the quality of the approximations with the JSD.

Note that while the KLD can take any non-negative values, the JSD is limited to the range $[0, \log(2)]$. This is true independently of the dimensionality of the distributions. So for completely discrepant $P$ and $Q$, we will have a value of $\log(2) \approx 0.693$.

\section{Auto-differentiation Implementation and GWDALI 1.0}

It is well established in the literature that computing numerical derivatives via finite differences is highly sensitive to the choice of step size. This problem becomes even more severe for higher-order derivatives, which can quickly become unreliable unless the step size is carefully tuned. Among the numerical methods available for computing derivatives, automatic-differentiation overcomes this limitation by evaluating derivatives up to machine precision~\cite{baydin2018automatic}.

The new release of \texttt{GWDALI} introduces the computation of derivatives for DALI tensors\footnote{The term `DALI tensors' refers to the multi-index coefficients of the DALI expansion in Eq.~\eqref{Eq:DALI}.} via automatic-differentiation, in addition to the previously implemented finite-difference strategies. As in earlier versions, \texttt{GWDALI 1.0} enables users to configure the detectors (including locations, orientations, arm opening angles, and sensitivities), select waveform models, and choose the parameter estimation method—either MCMC sampling or grid evaluation—for both exact and DALI likelihoods. The FM can also be computed either through MCMC sampling or by matrix inversion to obtain the covariance matrix. In this new version, users may select the derivative computation method, choosing between finite differences and automatic-differentiation. The latter relies on the recursive application of \texttt{JAX.grad()} to obtain higher-order derivatives.

Since \texttt{JAX.grad()} only operates on functions implemented within the \texttt{JAX} framework, it cannot be directly applied to waveform models provided by the python-C interface of \texttt{LAL}~\cite{swiglal}. To address this limitation, we have reimplemented the waveform approximants \texttt{TaylorF2}, \texttt{IMRPhenomA}, \texttt{IMRPhenomB}, \texttt{IMRPhenomC}, \texttt{IMRPhenomD}, and \texttt{IMRPhenomHM} within \texttt{JAX} \cite{phenomA,phenomB,phenomC,phenomD_p1,phenomD_p2,phenomHM}. These new implementations exhibit excellent agreement with those in \texttt{LAL}, while offering significantly improved computational performance through the just-in-time compilation capabilities of \texttt{JAX} (see Appendix \ref{app:waveforms}).

In certain cases, DALI may fail to provide accurate results for specific parameters, even when computing tensors through automatic-differentiation. However, these limitations can often be mitigated through suitable reparametrizations. To support this, the new release allows users to select among different parameterizations for distance, masses, and spins. A detailed analysis of the impact of such reparametrizations on DALI’s performance will be presented in the following sections.

\subsection{Smart choices of parameters}

As is well known, a Gaussian posterior follows from a likelihood which is Gaussian in the data, and model parameters which are linear in the data. Therefore, a non-Gaussian posterior could in principle be made Gaussian, or at least more Gaussian, by a suitable change of parameters. In practice, finding a suitable set of parameters is non-trivial, and simple recipes to Gaussianize likelihoods often fail to produce significant improvements~\cite{Wolz:2012sr}. 

Recently, a systematic analysis of this problem in the context of GW was performed in~\cite{Roulet:2022kot}. Here, we will refrain from this more systematic reparametrization. Instead, we focus mostly on the choice of two parameters: one describing the distance parameter, the other the second mass degree-of-freedom, besides the tightly constrained chirp mass. A smart choice of these parameters produce important changes in the posteriors. In the DALI approach there are further considerations on choosing the optimal parameters than those made in~\cite{Roulet:2022kot}. We will thus leave a more systematic exploration of this issue for future works.

Since the amplitude of the dominant mode in the GW is proportional to $1/d_L$, if one uses $d_L$ as a parameter the DALI expansion series will have oscillating signs on the derivatives with respect to $d_L$. It is therefore more sensible to use instead $1/d_L$ as a distance parameter. This was first realized in~\cite{Wang:2022kia}. In practice, we found that naïvely using  $d_L$ as a parameter yielded results for the doublet which were worse than those of the simple FM, and often the doublet produced a spurious second peak in the distance posterior, which in many cases was higher than the fiducial (injected) distance. This does not occur if one goes to the triplet, but this incurs extra computational cost. Using $1/d_L$ instead makes the doublet well-behaved, and results in similar accuracy than the full triplet, as we will show below.\footnote{One may also consider alternative parametrizations in order to test their performance against exact results.  Rather than recomputing the full set of DALI tensors, tensors in different $d_L$-parametrizations can be obtained through the transformation rules described in Appendix~\ref{app:tns_transf}. We tested using $d_L^{-2}$ and $d_L^{-1/2}$ and got comparable results to those obtained with $1/d_L$.}

Besides the distance parameter, another parameter that resulted in poor performance of DALI was the symmetric mass ratio $\eta \equiv m_1 m_2 / (m_1 + m_2)^2$. This parameter has, by definition, an upper bound at 0.25. Moreover, the observed population of BBH is strongly concentrated in the region around the $\eta > 0.24$ region. An expansion of the likelihood close to a parameter bound may result in an incorrect approximation, and this is what we found when using $\eta$. One could replace $\eta$ by the total mass $M \equiv m_1 + m_2$, but we found that using $\delta_M \equiv (m_1 - m_2) / M = \sqrt{1-4\eta}$ resulted in slightly more accurate results (see Appendix \ref{app:eta_vs_dM}). Even though it is also bound by zero, the distribution of injections is more spread out, and the derivatives are smoother. We therefore use as mass parameters the pair $\{{\cal M}, \delta_M \}$ being ${\cal M}$ the chirp mass. The generated samples can then be converted \emph{a posteriori} from $\{{\cal M}, \delta_M \}$ to $\{{\cal M}, \eta\}$ in order to present results for $\eta$.
For the spins, we use the symmetric and antisymmetric spin parameters $\chi_s \equiv (\chi_1 + \chi_2)/2$ and $\chi_a \equiv (\chi_1 - \chi_2)/2$.

\begin{figure}
\centering
\includegraphics[width=.85\columnwidth]{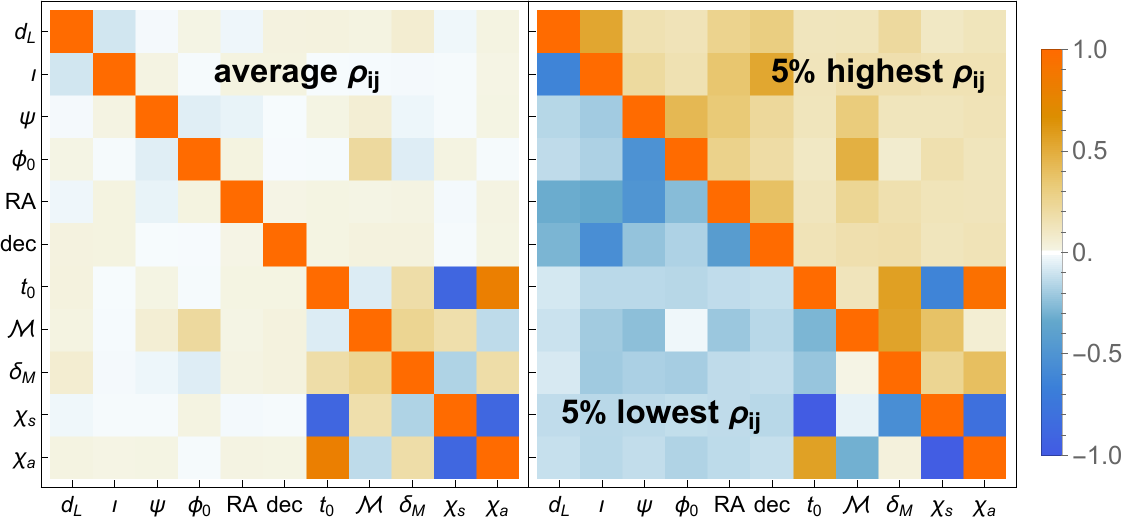}
    \caption{Correlation matrix $\rho_{ij}$ for the 11 parameters and the 300 GW detections here considered, using the exact posteriors. We show the average $\rho_{ij}$ coefficients (left panel), as well as the 5\% lowest  and 5\% highest (right panel) values for each coefficient. As it is well-known, the pair $d_L$, $\iota$ can be highly correlated or anti-correlated, and the pair $\chi_s$, $\chi_a$ are always highly anti-correlated. }
\label{fig:corr-matrix}
\end{figure}

In Figure~\ref{fig:corr-matrix} we show results for the correlation matrices $\rho_{ij}$ among the 11 parameters in the case of exact likelihoods for the 300 injections we have performed. For each of the correlation coefficients we can build histograms. We depict both the average, as well as the $5\%$ lowest and highest values. As it is well-known, the pair $d_L$, $\iota$ can be highly correlated or anti-correlated, even though the average correlation value is only slightly negative. We also find that the pair $\chi_s$, $\chi_a$ are always highly anti-correlated, with an average anti-correlation of $-0.89$. This could be mitigated in the future by using the effective spin parameter $\chi_{\rm eff}$, which is less correlated with $\chi_a$, and $\chi_a$ itself could be replaced by its mass-weighted counterpart. The spins are also found to be highly correlated with the coalescence time $t_0$.

\section{The MCMC Fisher hybrid and the issue of GW priors}\label{sec:MCMC-Fisher}

In most cases, in GW, except perhaps for events with both very high signal-to-noise ratio and measurements in multiple detectors, the FM results extend to unphysical values, such as angles well outside their $[0, \pi]$ or $[0, 2\pi]$ support. To prevent the FM from covering these unphysical regions of the parameter space, one can introduce priors to the analysis. 

If one wants to remain completely inside the Fisher formalism, the priors must be Gaussian functions, which are used to build a prior FM, which is then added to the likelihood FM. Adding Gaussian priors can improve substantially the condition number of the resulting posterior FM, and therefore its reliability. Therefore, in GW FM analysis one should always add at least very conservative Gaussian priors for the parameters with bounded support. See~\cite{Rodriguez:2013mla} for further discussion on prior effects on GW FM.  We tested the following simple, mildly informative priors, and we find them to substantially improve the FM performance:\footnote{Throughout this work, the parameters $\{d_L,\iota,\psi,\phi_0,{\rm RA},{\rm Dec},t_0, {\cal M}\}$ are measured in units of \{Gpc, rad, rad, rad, deg, deg, sec, $M_{\odot}$\} respectively, while \{$\delta_M,\chi_s,\chi_a$\} are dimensionless.}
\begin{equation}\label{eq:gauss-prior}
    F_{\rm prior} = {\rm diag}(\ \underset{ \atop d_L}{0}\ ,\ \underset{\atop\iota}{\pi}^{-2}\ ,\ \underset{\atop\psi}{\pi}^{-2}\ ,\ \underset{\atop \phi_0}{\pi}^{-2}\ ,\ \underset{\atop  \rm RA}{360}^{-2}\ ,\ \underset{\atop \rm Dec}{180}^{-2}\ ,\ \underset{\atop t_0}{10^{-2}} \ ,\ \underset{\atop {\cal M}}{100^{-2}} \ ,\ \underset{\atop \delta_M}{1}\ ,\ \underset{\atop \chi_s}{1}\ ,\ \underset{\atop \chi_a}{1}\ )\,.
\end{equation}
$F_{\rm prior}$ is only mildly informative in the sense that it assumes uncorrelated standard deviations for all parameters and simply mimic the total support range of the angular, $\delta_M$ and spin variables.\footnote{Another important effect of including these priors is the reduction in rejected sources, from 31\% down to only 0.7\% according to our inversion threshold condition.} Finally, following \cite{Iacovelli:2022bbs}, we deal only with sources with an inversion threshold $\epsilon_{\rm inv}$ are less than $0.05$, being $\epsilon_{\rm inv}$ defined below:
\begin{equation}
    \epsilon_{\rm inv}\equiv \underset{ij}{\rm max} \ |(F^{-1}\cdot F-\mathbb{1})_{ij}|\ .
\end{equation}
We tested smaller thresholds as alternatives,  but the results were similar.

\begin{table}
    \centering
    \begin{tabular}{ccc}
        \hline
        Parameter & Prior & Limits \\
        \hline
        $d_L$ &  Eqs.~\eqref{Eq:MD1}-\eqref{Eq:MD2} & $z\in[0,10]$   \\
        $\iota$ & Sine & $(0,\ \pi)$ \\
        $\psi$ & Uniform & $(0,\ \pi)$ \\
        $\phi_0$ & Uniform & $(0,\ \pi)$ \\
        ${\rm RA}$ & Uniform & $(-180^o,\ 180^o)$ \\
        ${\rm Dec}$ & Cosine & $(-90^o,\ 90^o)$ \\
        $t_0$ & Uniform & $(\hat{t}_0\pm100/\sqrt{F_{t_0,t_0}})$ \\
        $\mathcal{M}$ & Uniform & $(\hat{\mathcal{M}}\pm100/\sqrt{F_{\mathcal{M,M}}})$ \\
        $\delta_M$ & Linear & $(0,\ 1)$ \\
        $\chi_s$ & Uniform & $(-1,\ 1)$ \\
        $\chi_a$ & Uniform & $(-1,\ 1)$ \\
        \hline
    \end{tabular}
    \caption{Priors used for the exact, DALI and MCMC Fisher likelihoods. The hats denote injection values while $F_{i,j}$, are the $ij$ components of the FM.  For the prior in $1/d_L$ we use ${\rm prior}(1/d_L)={\rm prior}(d_L)\times d_L^2$. The linear prior in $\delta_M$ corresponds to a uniform prior in $\eta$.}
    \label{tab:Priors}
\end{table}

\begin{figure}
    \centering
    \includegraphics[trim=0mm 21.5mm 0mm 0mm, clip, width=.42\linewidth]{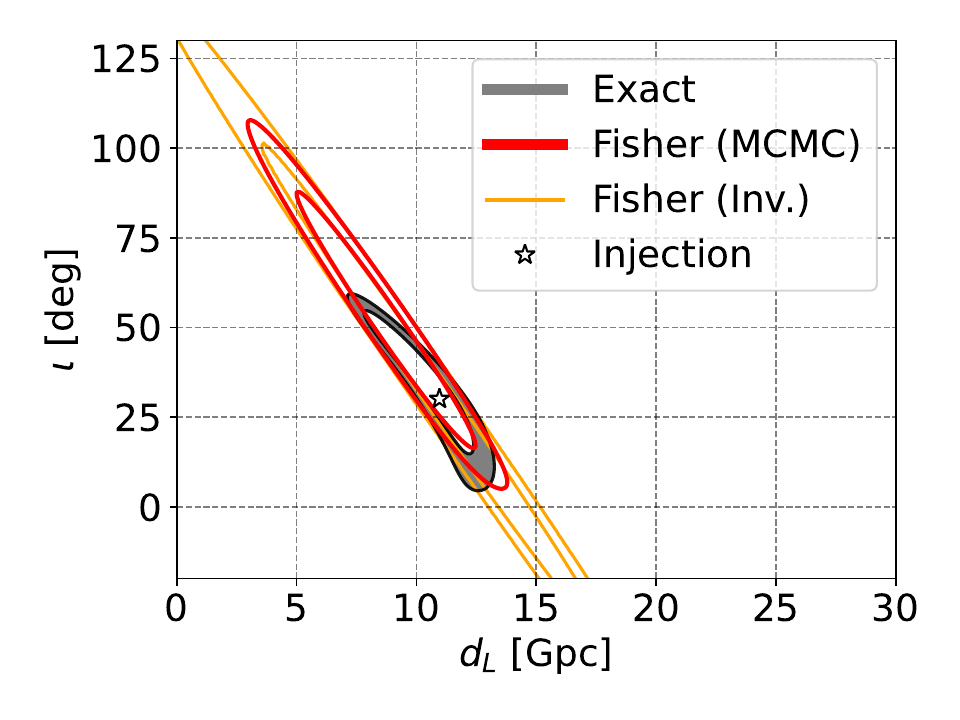}
    \includegraphics[trim=0mm 21.5mm 0mm 0mm, clip, width=.42\linewidth]{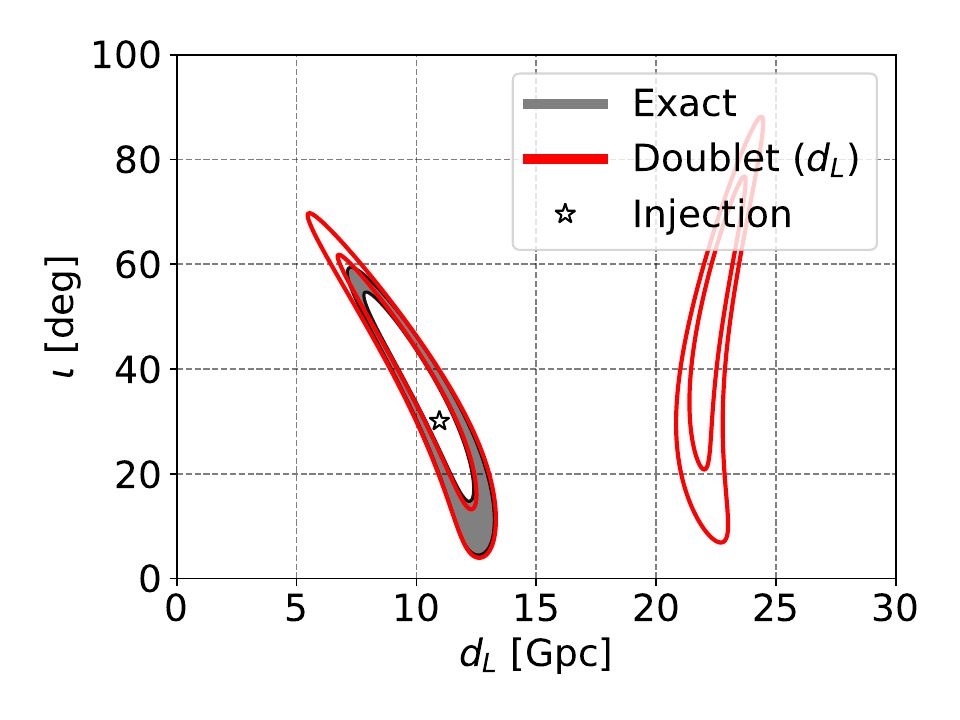}
    \includegraphics[width=.42\linewidth]{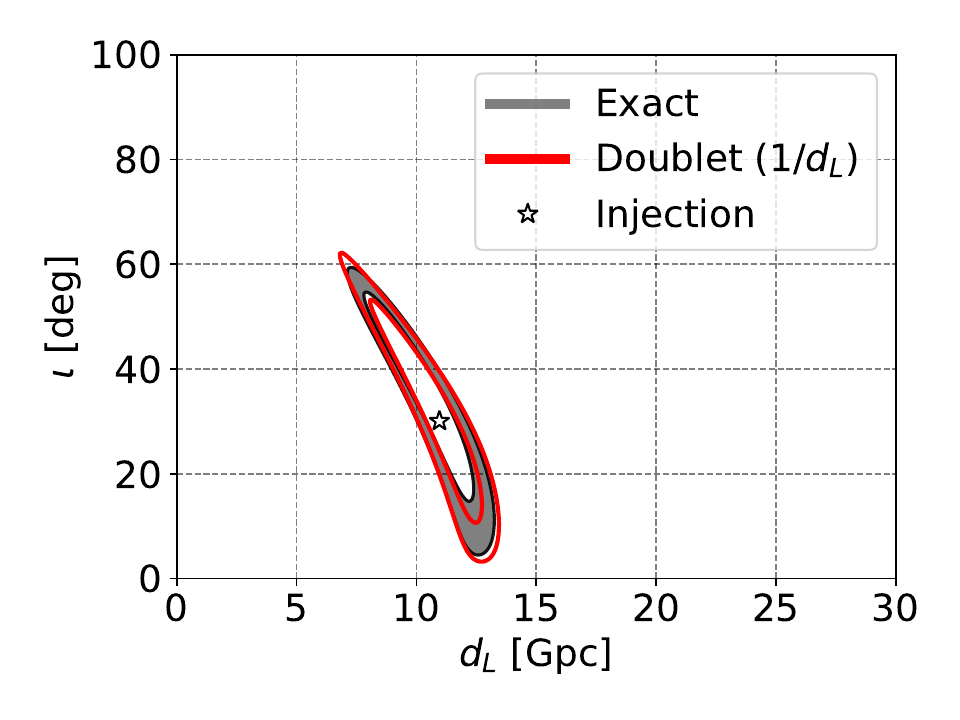}
    \includegraphics[width=.42\linewidth]{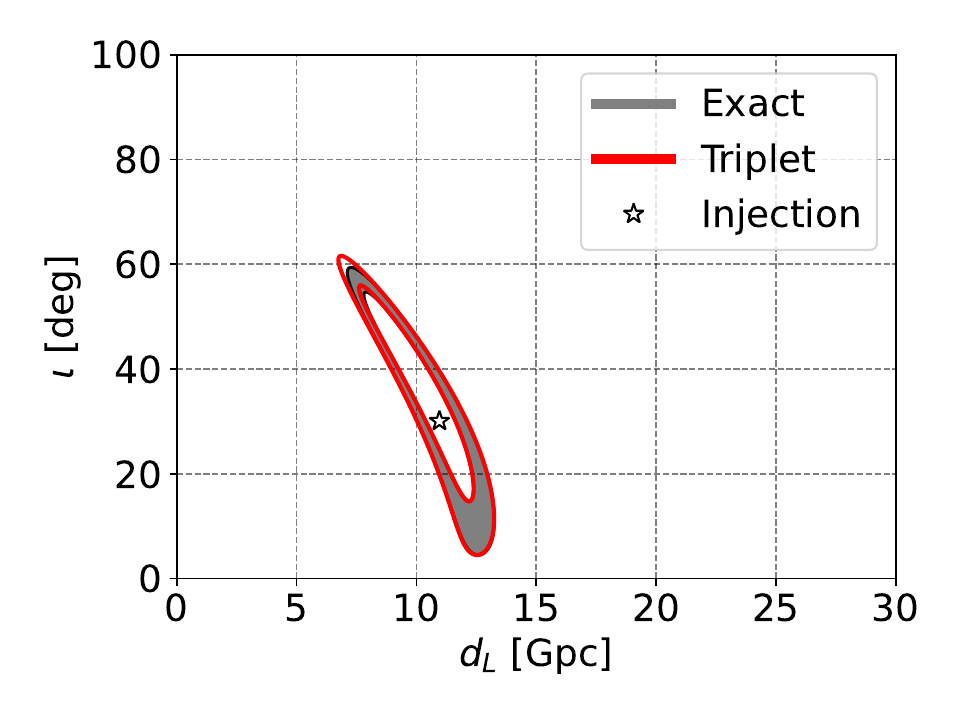}
    \caption{Two-dimensional $d_L-\iota$ posterior for a particular GW, comparing the exact result (black shaded) with the standard Fisher with Gaussian priors, the MCMC Fisher with exact priors and the DALI approximations. The source parameters used were: $d_L=10.96\,$Gpc, $\iota=30^o$, $\psi=133^o$, $\phi_{coal}=-78.7^o$, $\rm{RA}=-135.8^o$, $\rm{Dec}=72.9^o$, $t_{coal}=-0.53\ s$, $\mathcal{M}=9.6\ M_{\odot}$, $\eta=0.25$, $\chi_1=0.31$, $\chi_2=0.16$.}
    \label{fig:2D_exact_dali}
\end{figure}

Being constrained completely to the standard FM formalism imposes two important drawbacks, which are especially important in GW analysis. First, the priors have to be Gaussians, whereas for instance it would be more reasonable to use top-hat priors for all angular variables. Second, as discussed before, the FM inversion is often unreliable. An interesting alternative is to use the FM to build an exact Gaussian likelihood, which can be subsequently sampled using conventional MCMC methods or similar alternatives~\cite{Iacovelli:2022bbs,desouza:2023ozp,Dupletsa:2024gfl}. This removes both disadvantages above. Since both likelihood and priors are analytical functions, such sampling procedure is very fast and efficient.

In order to illustrate this issue, we can consider a much simplified parameter estimation in which we only vary the distance $d_L$ and the inclination $\iota$, which are known to have an important partial degeneracy. Figure~\ref{fig:2D_exact_dali} depicts the posteriors obtained for a particular, but typical, GW injection. In this plot and in what follows, we adopt the priors described in Table~\ref{tab:Priors}, which are much more realistic than the one described in Eq.~\eqref{eq:gauss-prior}. The advantage of restricting to this two-parameter case is that the FM inversion becomes very stable, and any difference between both FM methods is a result of the prior alone. We see that the traditional FM estimates using Gaussian priors and matrix inversion extends to negative $\iota$ values. This is improved with the full priors and MCMC sampling, but an accurate fit is only obtained with either the doublet or triplet-DALI methods. We also show the importance of using $1/d_L$ as distance parameter for the doublet-DALI case, which otherwise produces a very large spurious bi-modality.

\begin{figure}
    \centering
    \includegraphics[width=0.9\linewidth]{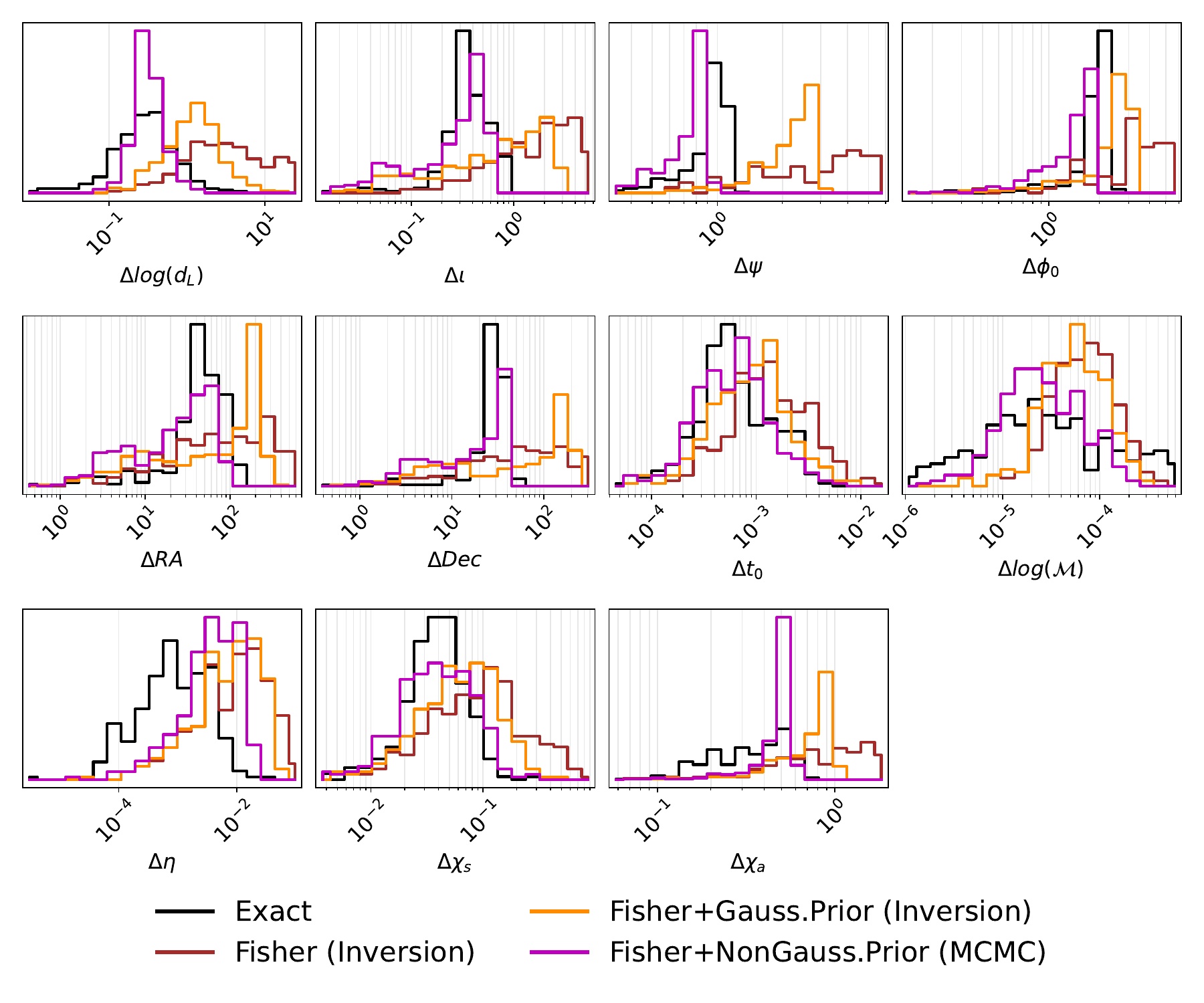}
    \caption{Distributions of parameter uncertainties obtained from Fisher. We compare three cases: using the inversion of the raw FM, using the inversion of FM + $F_{\rm prior}$ and through a MCMC sampling of the exact Gaussian likelihood with non-Gaussian priors.
    }
    \label{fig:FigFisherPriors}
\end{figure}

To better quantify how the different priors and FM schemes perform, we now revert to an 11-dimensional analysis of the GWs, and analyze a total of 300 sources with realistic parameter distribution and which could be observed by the Einstein Telescope. We describe these sources in more detail in the next section. Here we simply show in Figure~\ref{fig:FigFisherPriors} the results of using the raw FM, the FM summed with the Gaussian $F_{\rm prior}$, and an MCMC sampling of the FM with the priors in Table~\ref{tab:Priors}. Note that we assume uniform priors in $\chi_s$ and $\chi_a$, which is not equivalent from assuming uniform $\chi_1$, $\chi_2$ priors.

As can be clearly seen, simply adding a minimalistic prior strongly affects the results. This is not only true for the parameters for which $F_{\rm prior}$ is non-zero, but also to the distribution of other parameters due to the correlations between them, in particular to $d_L$  (which is highly correlated to $\iota$). However, foregoing the inversion altogether (which for the 11-parameter case is often no longer accurate) and adding realistic priors impose very significant changes to the results. As we will show next, this procedure is much more accurate when compared to full MCMC chains of the exact posterior.  Here we will refer to it as the \emph{MCMC Fisher}, but it could also be called the \emph{singlet-DALI} to emphasize that it follows the standard DALI procedure and not matrix inversions.

\section{DALI performance for the Einstein Telescope}\label{sec:ET}

Here we investigate how DALI performs for GWs in the particular case of the next-generation Einstein Telescope. We will assume throughout a triangular configuration for the Einstein Telescope with coordinates $lon=6^o$ and $lat=50^o$. We assume the ET-D sensitivity curve.\footnote{Sensitivity curve available in \href{https://apps.et-gw.eu/tds/?r=14065}{https://apps.et-gw.eu/tds/?r=14065}.}

We will limit ourselves to GWs from BBH mergers, and leave a more systematic study of BNS and NSBH, as well as other detector configurations, for future work. For the redshift distribution of BBH, we assume it to be proportional to the comoving volume as follows
\begin{equation}\label{Eq:MD1}
    P(z) = \frac{1}{\mathcal{N}}\frac{\psi_{\rm MD}(z)}{1+z}\frac{\dd V_c}{\dd z}   \,,
\end{equation}
where $\psi_{MD}(z)$ is the Madau-Dickinson distribution profile with the functional form~\cite{Madau:2014bja,Madau:2016jbv}:
\begin{equation}\label{Eq:MD2}
    \psi_{\rm MD}(z)\equiv \frac{(1+z)^a}{\left[1+\left(\frac{1+z}{1+z_p}\right)^{a+b}\right]} \, .
\end{equation}
For our BBH simulations, we set $a=2.7$, $b=3.0$, and $z_p=2.0$, following \cite{Iacovelli:2022bbs}, where these values were obtained by fitting BBH populations to the GWTC-3 catalogs. These parameters ($a, b, z_p$) differ slightly from those in the original Madau–Dickinson star-formation rate, as a time delay between the formation of stellar progenitors and the eventual mergers of BBH is generally expected, and this moves the sources to smaller redshifts~\cite{Vitale:2018yhm}. The conversion $z-d_L$ was done using a flat-$\Lambda$CDM cosmology with $H_0=70\,$km/s/Mpc and $\Omega_{m0}=0.3$.

We assume a BH mass distribution following the Power-Law + Peak model, which was the standard for GWTC-3, following \cite{Iacovelli:2022bbs}. For the distribution of the mass ratio $q$ we use the analytical fit for the GWTC-3 as follows \cite{MenoteMarra:pc}:
\begin{equation}
    P(q) = A_4\cdot\left[A_1\cdot q+ \frac{A_2}{q} + \frac{A_3}{q^2}\right]\,,
\end{equation}
with $A_1=7.98$, $A_2=1.46$, $A_3=-0.209$ and $A_4=0.01$. Finally, for the distribution of spins, we assume:
\begin{equation}
    P(\chi_{1,2}) \propto |\chi_{1,2}|^{a-1}\cdot (1-|\chi_{1,2}|)^{b-1}\,,
\end{equation}
where $a=1.2$ and $b=4.12$ following~\cite{Iacovelli:2022bbs}.

We assume in all cases the waveform approximant \texttt{IMRPhenomHM} \cite{phenomHM} which makes use of the  high harmonics modes (2,2), (2,1), (3,3), (3,2), (4,4) and (4,3) in the spin-weighted spherical harmonics decomposition of the strain up to the 3.5 PN:
    \begin{equation}
        h \equiv h_+-ih_{\times}=\sum_{4\geq \ell \geq2} \sum_{|m|\leq \ell}{}_{-2}\mathcal{Y}^{\ell m}(\iota,\phi_0)h_{\ell m}\,,
    \end{equation}
where ${}_{-2}\mathcal{Y}^{\ell m}$ are the spin-weighted spherical harmonics with spin-weight $s=-2$.

In this paper we will set a minimum SNR of 8 for all GW detections, which is the lowest value usually considered in the literature. This is the value used in all results below. Clearly, higher thresholds should increase the Gaussianity of the posteriors, and improve both the DALI and Fisher approximations. We will briefly quantify this in a few cases.

\subsection{Uncertainty histograms}

We now compare the marginalized standard deviations found for all 300 injections when we analyze them with either the exact, the FM, the doublet-DALI and the triplet-DALI. Often, when forecasting for future detectors, one is more concerned with how well we can forecast their precision, and the possibility of biases in the estimation is of secondary concern. So this is the simplest possible meaningful comparison between the likelihood method. In the case of the doublet-DALI, as discussed previously, we use $1/d_L$ as distance parameter. For exact, DALI and MCMC Fisher likelihoods, we use the priors described in Table~\ref{tab:Priors}. For the prior in $d_L$ we use the MD-profile with $a=2.7,\ b=2.9,\ z_p=2.9$ (as in \cite{Madau:2014bja}) from a flat-$\Lambda$CDM model with $H_0=70$\, km/s/Mpc and $\Omega_{m0}=0.3$.

\begin{figure}
\centering
\includegraphics[width=.85\columnwidth]{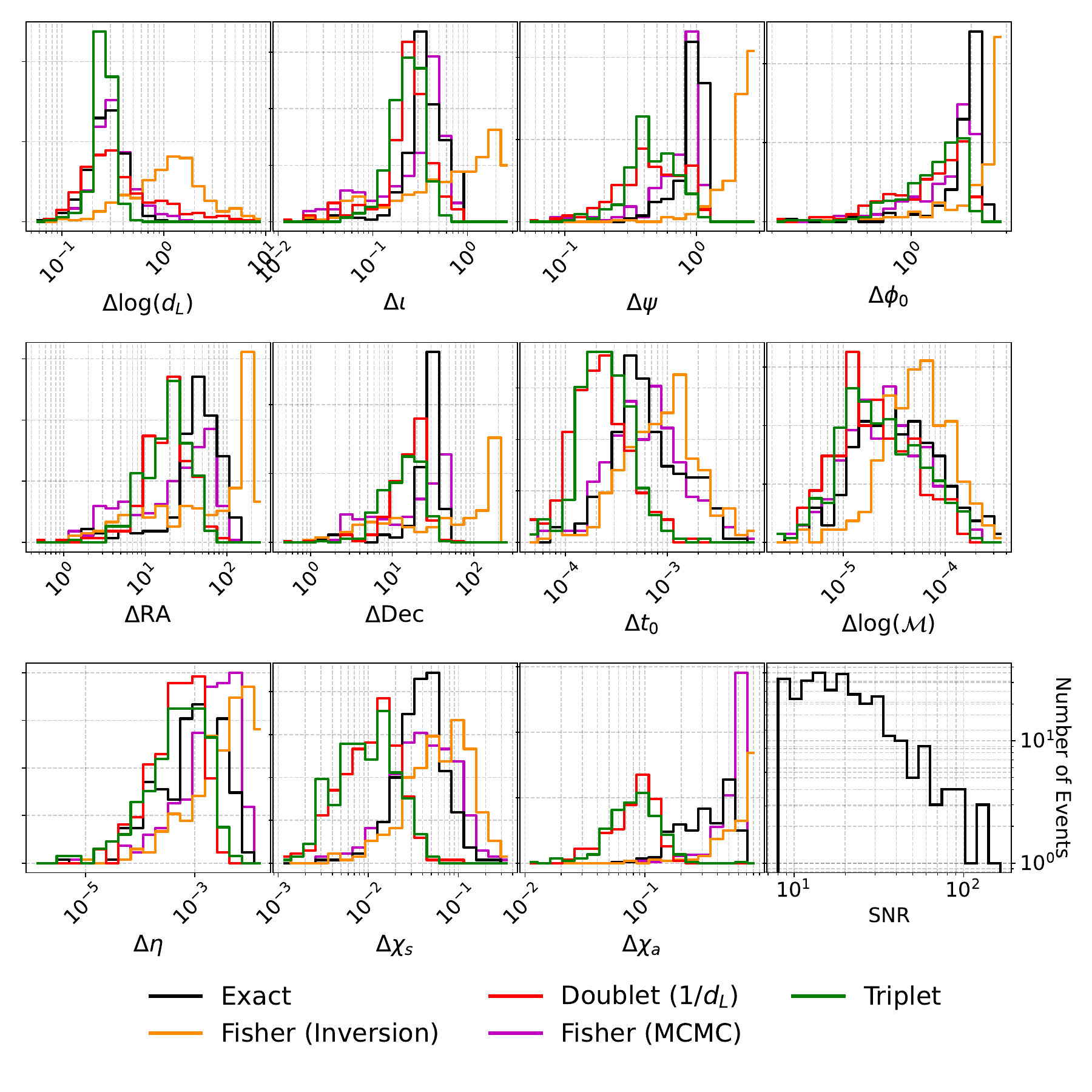} 
    \caption{
    Comparison of the standard deviation realizations for the marginalized posterior in each parameter. We see that the FM tends to overestimate the uncertainties, while the triplet yields the most accuracy. The doublet using $1/d_L$ is a good compromise in general. In addition, we show in the last panel the SNR distribution of the corresponding sources.
    }
\label{fig:error-histograms}
\end{figure}

Figure~\ref{fig:error-histograms} depicts these results. As can be seen, for all parameters on average the FM overestimates uncertainties by a large degree. This agrees with the findings in~\cite{Rodriguez:2013mla}. This is especially true for the luminosity distance. The DALI performs much better. The triplet case produces slightly better estimates than the doublet for parameters like $\eta$ and $\chi_a$. Both DALI approximations perform very well for $d_L$, but underestimate most other parameters. The discrepancy is larger for the spins and $\psi$, but smaller for the localization parameters. 

For GW cosmology, besides the straightforward bright siren method, two other ways in which one can reconstruct the expansion history of the universe which are often considered promising. The first is the spectral siren method, which relies on features of the GW mass function to infer the redshift of the sources and thus cosmology~\cite{Taylor:2011fs, Farr:2019twy, Mastrogiovanni:2021wsd, Mancarella:2021ecn, Mukherjee:2021rtw,Ezquiaga:2022zkx,Karathanasis:2022rtr}. The second relies on cross-correlation with galaxies, either on an individual basis (the line-of-sight dark sirens method)~\cite{DelPozzo:2011vcw, LIGOScientific:2018gmd, PhysRevD.101.122001}, or, as recently proposed, on the catalog-level (the peak sirens method)~\cite{Ferri:2024amc, SantiagodeMatos:2025iyj}. Reliable distance estimates are also crucial for forecasts of modified gravity test that rely on possible discrepancies between luminosity and GW distances~\cite{Nishizawa:2017nef, Amendola:2017ovw, Belgacem:2018lbp, Matos:2022uew, Matos:2023jkn}. In all cases, the performance depends on how well the GW distance can be inferred, and in the line-of-sight and peak sirens methods, also on a good localization estimation.

We conclude that for GW cosmology, DALI yields much more realistic marginalized uncertainty estimates than the classic FM with inversions. For the MCMC Fisher, the uncertainties have comparable distributions with those of both DALI methods, with the best results alternating depending on the chosen parameter.

\subsection{How accurate is DALI?}

\begin{figure}
    \centering
    \includegraphics[width=.83\linewidth]{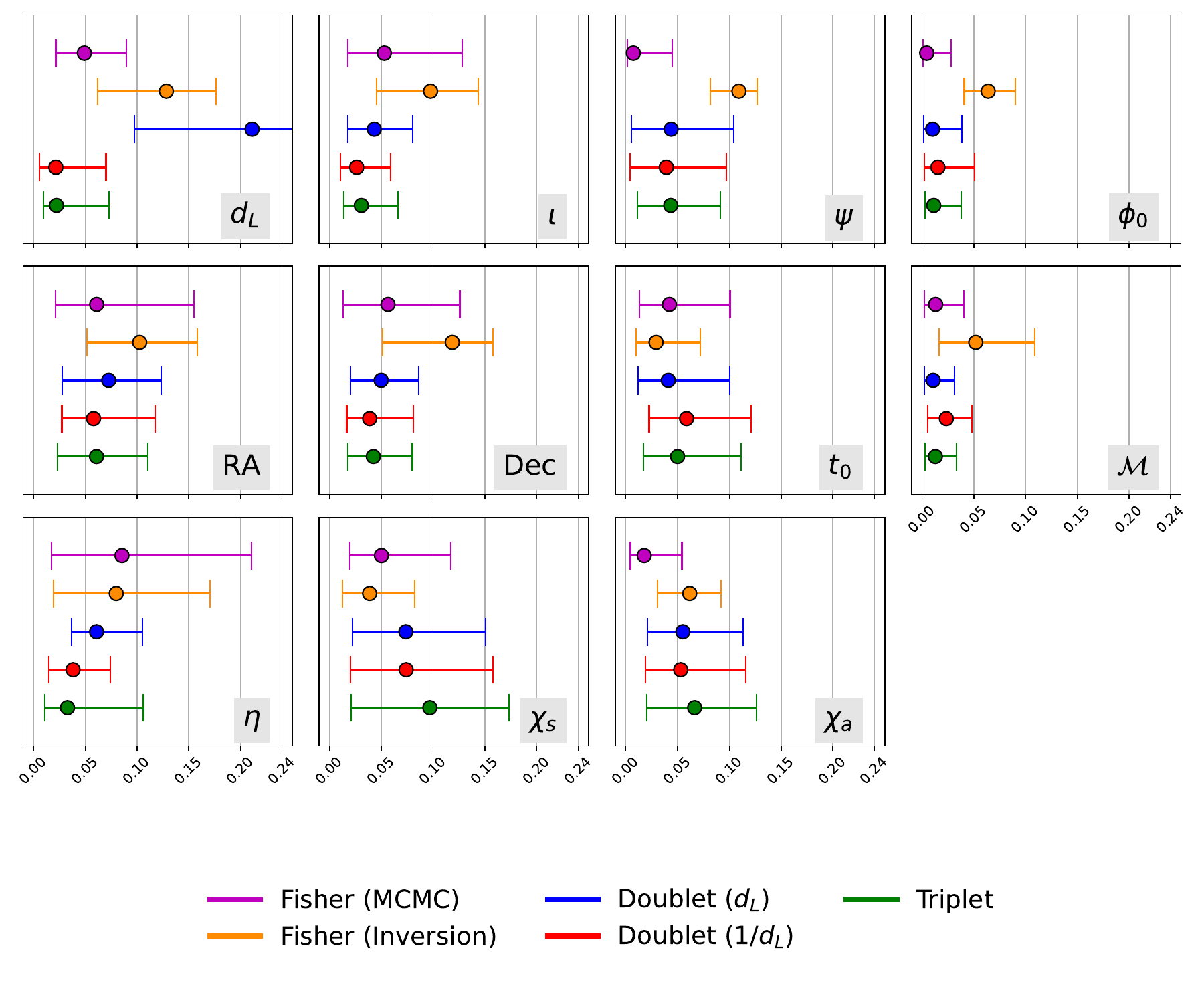}
    \caption{
    Median and 68\% quantiles 1-D Jenson-Shannon Divergences between the exact posterior and each of the approximations for the marginalized posteriors in each of the individual parameters for the 300 injections. We see that both FM inversion and the doublet DALI approximation using $d_L$ as parameter are the most inaccurate. We also find that the doublet-DALI with $1/d_L$ results in small JSD, comparable to the triplet using $d_L$.}
    \label{fig:JSD-1D}
\end{figure}

Clearly, marginalized 1-D histograms tell only a small part of the story. First, they contain no information on how biased the results are. Second, for some GW some parameters could have underestimated uncertainties while others have overestimated ones. For distance and localization, for instance, this could lead to an incorrect understanding of the potential in cross-correlating with galaxies. In order to better understand how well DALI performs, we will use the Jenson-Shannon Divergence, discussed above. 

We start by evaluating the JSD for the 1-D marginalized posteriors in each of the 11 GW parameters here considered. We compute the JSD for all 300 injections, comparing each approximation scheme with the exact MCMC posterior. The results are shown in Figure~\ref{fig:JSD-1D}, where we depict the median JSD values and the $68\%$ quantile range of the JSD distribution. Here we also include the naïve doublet case in which $d_L$ is used as a parameter instead of $1/d_L$. We find that DALI significantly outperforms the FM inversion for all parameters except $t_0$ and $\chi_s$. Both the doublet-DALI using $1/d_L$ and the triplet-DALI show comparable results, whereas the doublet using $d_L$ is the least accurate for the distance inference, as already discussed.

\begin{figure}
\centering
\includegraphics[width=.92\linewidth]{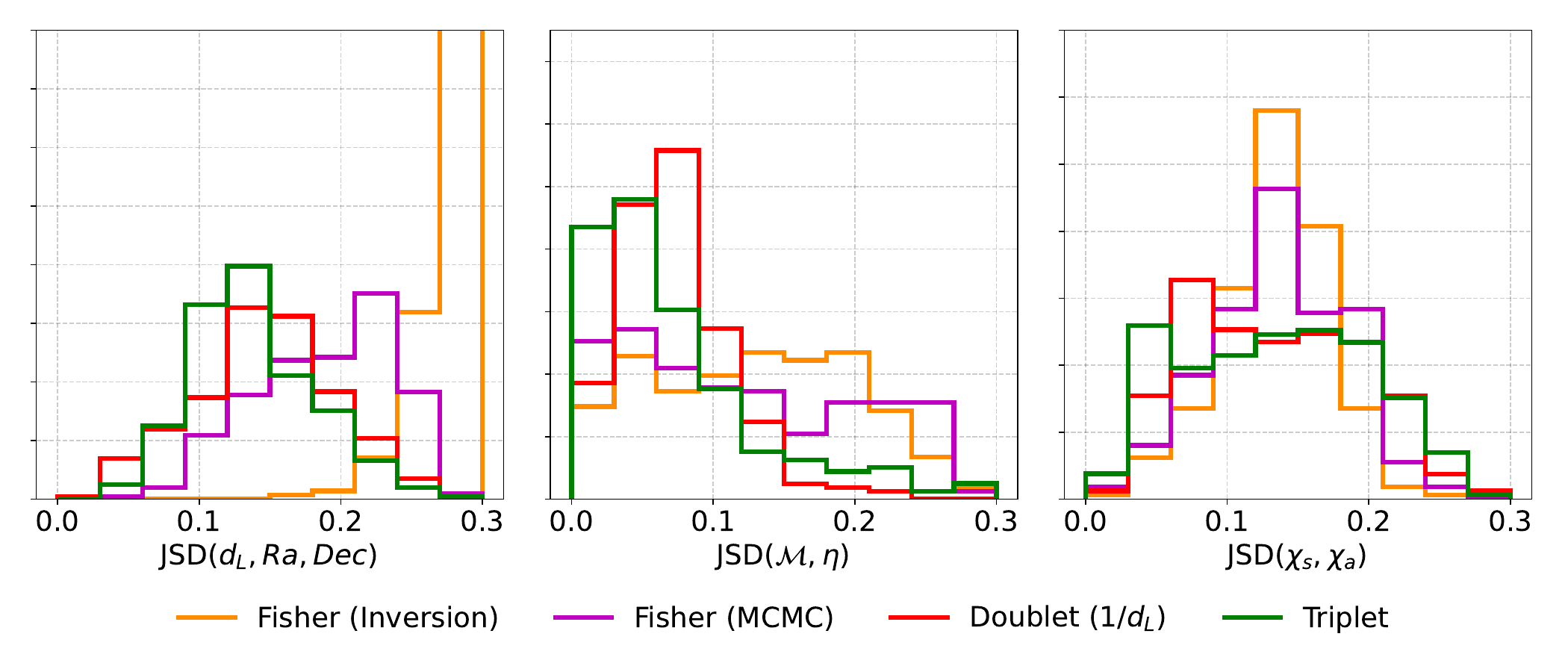}
    \caption{
    Jenson-Shannon Divergences for the 3-D posterior $\{d_L, {\rm RA}, {\rm Dec}\}$ (left), and for the 2-D posteriors $\{{\cal M}, \eta\}$ (middle) and  $\{\chi_s, \chi_a\}$ (right panel). DALI is clearly more accurate than Fisher, and using $1/d_L$ as a parameter for the doublet yields large improvements for the $d_L$ posteriors.
    }
\label{fig:JSD-3D}
\end{figure}

As pointed out above, for the dark siren method one relies on good estimates of the volume in space where the GW occurred. So it is also useful to compute the JSD not only for the marginalized 1-D posteriors, but on the full 3-D posterior on $\{d_L, {\rm RA}, {\rm Dec}\}$. Figure~\ref{fig:JSD-3D} depicts the 3-D JSD histogram for $\{d_L, {\rm RA}, {\rm Dec}\}$. We also show the results for the 2-D JSD for the mass parameters $\{{\cal M}, \eta\}$ and for the 2-D spin parameters $\{\chi_s, \chi_a\}$. 

Table~\ref{tab:jsd_values} summarizes our JSD results for each individual parameter, as well as for the 3-D volume posteriors, the 2-D mass posteriors and the 2-D spin posteriors. For the 1-D marginalized posteriors, the simple MCMC Fisher method (singlet-DALI) already provides a reasonable approximation, with results comparable to the doublet-DALI and triplet-DALI. The DALI benefits, however, become clear when considering the posterior in higher dimensions. Both the 3-D volume posteriors and 2-D mass posteriors are significantly more accurately approximated by the doublet-DALI than the MCMC Fisher. And the Fisher with simple inversions, as discussed above, is very inaccurate. Table~\ref{tab:jsd_values} also shows that GW with larger SNR tend to be better approximated by Fisher or DALI, as expected. For the 1-D posteriors, the average JSD among the 11 parameters improves by a modest $\sim 10\%$ when we change from our default minimum SNR threshold of 8 to 20. Therefore we conclude that one would need really high SNR in order for the simple Fisher approximation to improve by a significant amount.

In Appendix~\ref{app:corner} we illustrate a corner plot, for a typical GW source, for all 11 parameters, comparing the exact MCMC contours with those obtained with FM, doublet-DALI and triplet-DALI.

\setlength\tabcolsep{4.0pt}
\begin{table}[t]
    \centering
    \small
	\begin{tabular}{lcccccccc}
		\toprule
		parameter set & \multicolumn{2}{c}{Fisher (Inversion)} & \multicolumn{2}{c}{MCMC Fisher} & \multicolumn{2}{c}{Doublet ($d_L^{-1}$)}  & \multicolumn{2}{c}{Triplet} \\
         & ${\rm SNR}\!\ge\!8$ & ${\rm SNR}\!\ge\!20$ & ${\rm SNR}\!\ge\!8$ & ${\rm SNR}\!\ge\!20$ & ${\rm SNR}\!\ge\!8$ & ${\rm SNR}\!\ge\!20$ & ${\rm SNR}\!\ge\!8$ & ${\rm SNR}\!\ge\!20$\\
		\midrule
        $d_L,\rm{RA},\rm{Dec}$ & 0.293 & 0.300 & 0.194 & 0.202 & 0.147 & 0.143 & 0.134 & 0.126\\
        $\mathcal{M},\eta$ & 0.135 & 0.090 & 0.105 & 0.065 & 0.068 & 0.054 & 0.052 & 0.032\\
        $\chi_s,\chi_a$ & 0.135 & 0.123 & 0.138 & 0.132 & 0.131 & 0.128 & 0.138 & 0.133\\
        \hline
        1-D average & 0.080 & 0.070 & 0.040 & 0.037 & 0.041 & 0.040 & 0.043 & 0.038\\
		\bottomrule
	\end{tabular} 
    \caption{Median of Jensen-Shannon Divergence (JSD) values for all sources for the different posterior approximations, and for two different SNR ratio thresholds, to wit 8 or 20. The last row shows the average marginalized 1-D JSD over all 11 parameters. The 1-D  posteriors are well-captured even with the MCMC Fisher method, but DALI performs much better in the 3-D volume posterior, and in the 2-D mass posteriors. Events with the larger minimum SNR tend to have only slightly lower JSD values.
    }
    \label{tab:jsd_values}
\end{table}

\subsection{Computational Cost Savings}

\begin{figure}
\centering
\includegraphics[width=.5\linewidth]{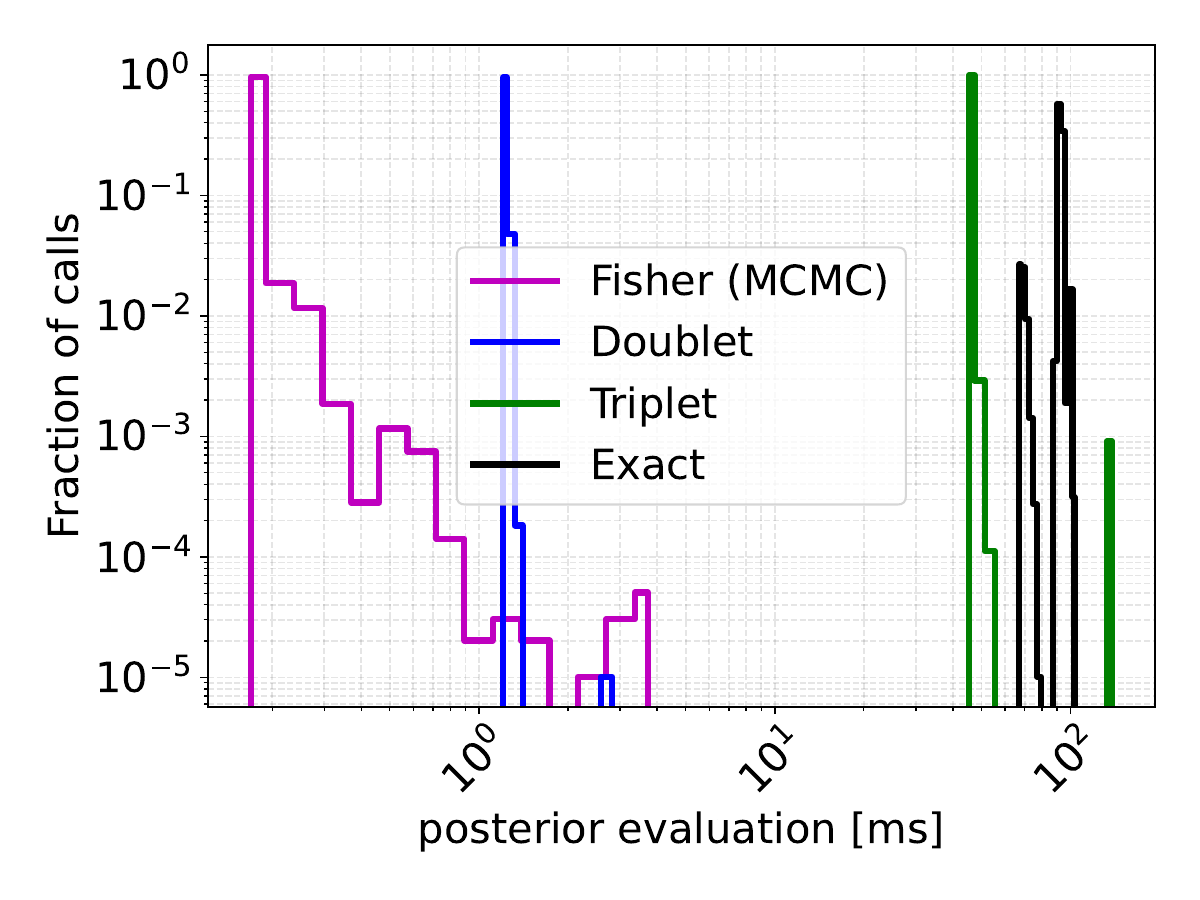}
    \caption{Single-thread computational time for each posterior evaluation for the exact posterior and the DALI approximations, assuming the DALI tensors were pre-computed. We combine times from 1000 evaluations from each of 100 different sources. The mean time are 0.2 ms (MCMC Fisher), 1.2 ms (doublet), 44.5 ms (triplet) e 91.1 ms (exact)
    }
\label{fig:cputime-histograms}
\end{figure}

The main advantage of using DALI is to reduce the computational cost of the likelihood evaluations. For each source, we first precompute all terms of the DALI tensors, such as $\langle \partial_i h | \partial_j\partial_k h \rangle$,  $\langle \partial_i\partial_j h | \partial_k\partial_l h \rangle$, etc. This procedure takes some time. But the subsequent steps, to wit the likelihood evaluations at each step of the MCMC chain, become much faster. To show these benefits in practice, we selected 100 sources from our mock catalog and computed 1000 likelihood evaluations for each, without any computational parallelization. Figure~\ref{fig:cputime-histograms} shows the results for doublet-DALI, triplet-DALI, and exact likelihoods.\footnote{ We note that the exact source of the multimodality in the computational cost, seen in Figure ~\ref{fig:cputime-histograms}, is still under investigation even after several tests.} We find that the doublet-DALI is almost two orders of magnitude faster than the exact likelihood, with a median computational time ($t_{\rm eval}$) of only 1.2 ms, compared to 91.1 ms for the exact case. The triplet case is, on the other hand, only two times faster. This means that here the doublet is the best approximation in terms of cost-benefit.

For a typical determination of a likelihood in 11 dimensions, one often needs to compute tens of thousands of evaluations in order to achieve convergence. This clearly depends on the technique used. GWDALI is fully modular, and allows the use of different samplers. Here we employ the \texttt{ptemcee} code~\cite{Vousden:2016eeu}, which is an MCMC approach that uses ensemble sampling with parallel tempering, as implemented inside the \texttt{bilby} library~\cite{ashton:2018jfp}. For this dimensionality \texttt{ptemcee} is much more efficient than standard ensemble sampling~\cite{Albert:2024zsh}. We employ 34 walkers and 6 temperatures in all cases. The convergence of MCMC chains depends not only on the chosen algorithm, but also on the shape of the posterior. Multimodality and strong correlations can both make it harder for chains to reach convergence. For the present case, we use two convergence criteria: a small Gelman-Rubin score ($\hat{R}-1<0.02$)~\cite{Gelman:1992zz} and the need for 50000 effective samples among all walkers, where by effective we mean the number of samples after discarding the burn-in and dividing by the chain autocorrelation. For the exact posteriors, convergence took on average  30000 steps for each walker and temperature.

Concerning our present investigation, we are interested in the relative gains when using DALI, and not in the absolute time to run the MCMC chains, which will depend on the chosen sampling code. We note that since DALI is a perturbative expansion for which the leading order is a Gaussian distribution, it can also result in faster convergence of the chains. We found that on average 55\% less steps were needed for the MCMC Fisher but only 40\% less for doublet-DALI.\footnote{Using the doublet-DALI with $d_L$ as a distance parameter instead of $1/d_L$ actually results in the need for more steps and not less, due to the frequent appearance of a spurious $d_L$ bimodality.} Nevertheless, most of the speed-up in the end still comes from the faster likelihood evaluations discussed above.

Finally, as mentioned above, for the DALI implementation there is also the computational cost ($t_{\rm pre}$) of pre-computing all tensors. In \texttt{GWDALI 1.0}, using the \texttt{IMRPhenomHM} waveform and auto-diff, for each source this takes on average 0.08 (1.1) [8.9] hours for the FM (doublet) [triplet] case. This means that for each source, a complete evaluation of the DALI takes a total average time $t_{\rm tot}$ given by
\begin{equation}
    t_{\rm tot} = t_{\rm pre} + t_{\rm eval} N_{\rm steps} N_{\rm walkers} N_{\rm temps}\,.
\end{equation}
For the MCMC Fisher (doublet) [triplet], this amounts to $0.19$h ($1.97$h) [$46.1$h], which should be compared to $109$h for the exact case. Therefore, using the doublet-DALI amounts to a total computational cost savings of a factor of 55 over the exact likelihood when using~\texttt{ptemcee}. This factor increases (decreases) if one uses a more strict (less strict) convergence criterion, requiring more (less) MCMC steps. The MCMC Fisher is also a reasonable candidate if faster evaluations are needed. It is over 500 times faster than the exact MCMC, and provides a much better cost-benefit than the simple FM inversion, since it takes less than triple the computational time and is much more accurate.\footnote{As the MCMC Fisher method is equivalent to re-weighting Gaussian samples according to the prior, more efficient implementations are possible -- see~\cite{Dupletsa:2024gfl}.} Similar conclusions for the MCMC Fisher were drawn in~\cite{Dupletsa:2024gfl}.

\section{Conclusions and Perspectives}

We have shown that the DALI approximation, and in particular the doublet-DALI, can be very useful to get estimations of the posteriors which are much faster than traditional MCMC methods, and much more accurate than standard Fisher Matrix methods.  This will allow for more accurate forecasts of the capabilities of detector upgrades and future detectors.

We have also made a more in-depth analysis of the hybrid MCMC Fisher method, in which the FM is used simply to compute an exact likelihood which is subsequently sampled through an MCMC (or similar) method. This allows for the use of exact priors, instead of simple Gaussian ones. We show that for 1-D marginalized constraints this method performs with similar accuracy than the doublet-DALI, but fails to capture higher dimensional posteriors with similar accuracy. Nevertheless, it is even faster, and {it turns out to be an interesting and fairly accurate} tool for fast evaluations. 

On the other hand, as currently implemented the triplet-DALI does not lead to significant improvements in terms of accuracy, and its larger computational time makes it an inconvenient trade-off to a standard MCMC.

Fast posterior evaluation can be important beyond forecasting a large number of future GW observations. It can also be useful to get approximate distributions of real data as quickly as possible and thus helpful for follow-up telescope triggering in the hopes of catching any related electromagnetic emission in its early stages.

There are a number of ways in which our approach can be further improved. First, the addition of other detectors to the network should enhance the localization of the sources, which should make it more Gaussian and thus better approximated by DALI. Second, it is possible to improve Gaussianity and reduce multi-modality by using alternative parametrizations for the GW, again resulting in more accurate DALI approximations. Third, our implementation of auto-differentiation using \texttt{JAX} is shown to be reliable, but at the same time the computation of higher derivatives still takes a meaningful fraction of the total computational time. It is possible that alternative methods, for instance using Julia as implemented in~\cite{Begnoni:2025oyd}, could result in faster derivative calculations. It is also possible to use the implemented auto-diff method to calibrate standard numerical derivatives and test if similar accuracy could be maintained, resulting in a further increase in efficiency.

We noted that the use of the doublet or triplet-DALI leads to an underestimation of uncertainties in some parameters, in particular on $\psi$, $t_0$, $\chi_s$ and $\chi_a$, and to a lesser degree also RA and Dec. Visual inspection of 300 corner plots similar to Figure~\ref{fig:corner-example-1} in Appendix~\ref{app:corner} indicates that the reason is that the posterior for parameters like RA, Dec and $\psi$ are often  multimodal. In these cases, DALI captures well the main peak, but is often incapable of reproducing the secondary peak. This leads to an underestimation of these parameters. As discussed above, a convenient reparametrization could potentially mitigate this issue.  For RA and Dec in particular, preliminary tests show that the inclusion of additional detectors improves substantially the DALI performance by suppressing this muldimodality. We plan to extend these investigations in  a future work.

We have limited ourselves here to the \texttt{IMRPhenomHM} waveform, which is one of the simplest waveforms which includes the higher-modes, which in turn are very relevant to get most of the distance information. We also assumed aligned spins. The field of waveforms is evolving fast, and further studies should be performed with different waveforms and more degrees of freedom in the parameter space.

\section*{Acknowledgments}
    The authors thank Davi Rodrigues, Felipe Barbosa, Isabela Matos, Ranier Menote and Valerio Marra for useful discussions. JMSdS is supported by the Brazilian research agency Fundação Carlos Chagas Filho de Amparo à Pesquisa do Estado do Rio de Janeiro (FAPERJ).
    MQ is supported by the Brazilian research agencies FAPERJ project E-26/201.237/2022, CNPq (Conselho Nacional de Desenvolvimento Científico e Tecnológico) and CAPES. We acknowledge the use of the computational resources of the joint CHE / Milliways cluster, supported by a FAPERJ grant E-26/210.130/2023.

\section*{Appendix}

\appendix

\section{Examples of GW corner plots using DALI}\label{app:corner}

\begin{figure}[t]
    \centering
    \includegraphics[width=\linewidth]{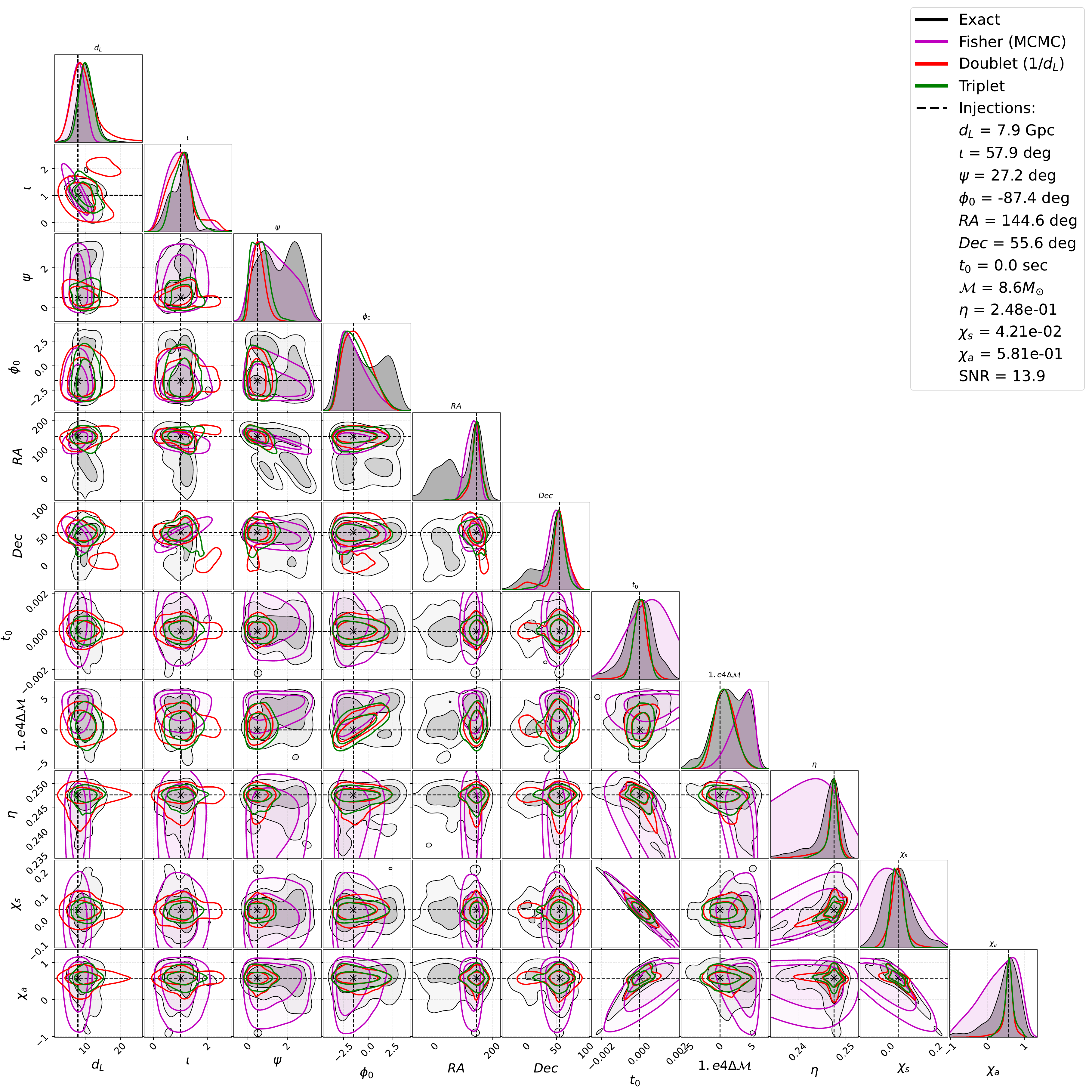}
    \caption{Corner plot from a typical source, obtained with the different posterior approximations. The injection parameters are shown in the plot legend.}
    \label{fig:corner-example-1}
\end{figure}

\begin{figure}[t]
    \centering
    \includegraphics[width=\linewidth]{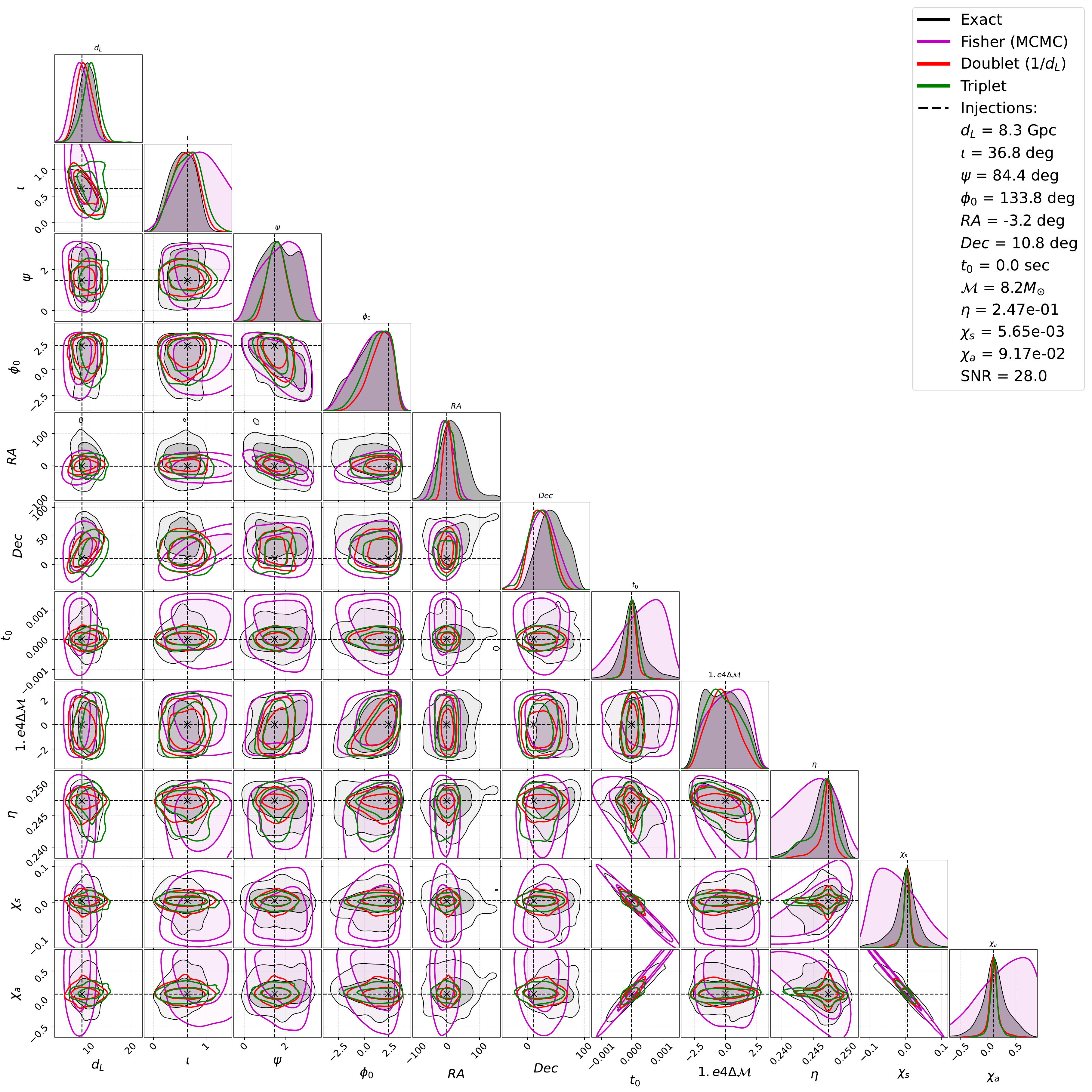}
    \caption{Same as Figure \ref{fig:corner-example-1} for a second, different source.}
    \label{fig:corner-example-2}
\end{figure}

\begin{figure}[t]
    \centering
    \includegraphics[width=\linewidth]{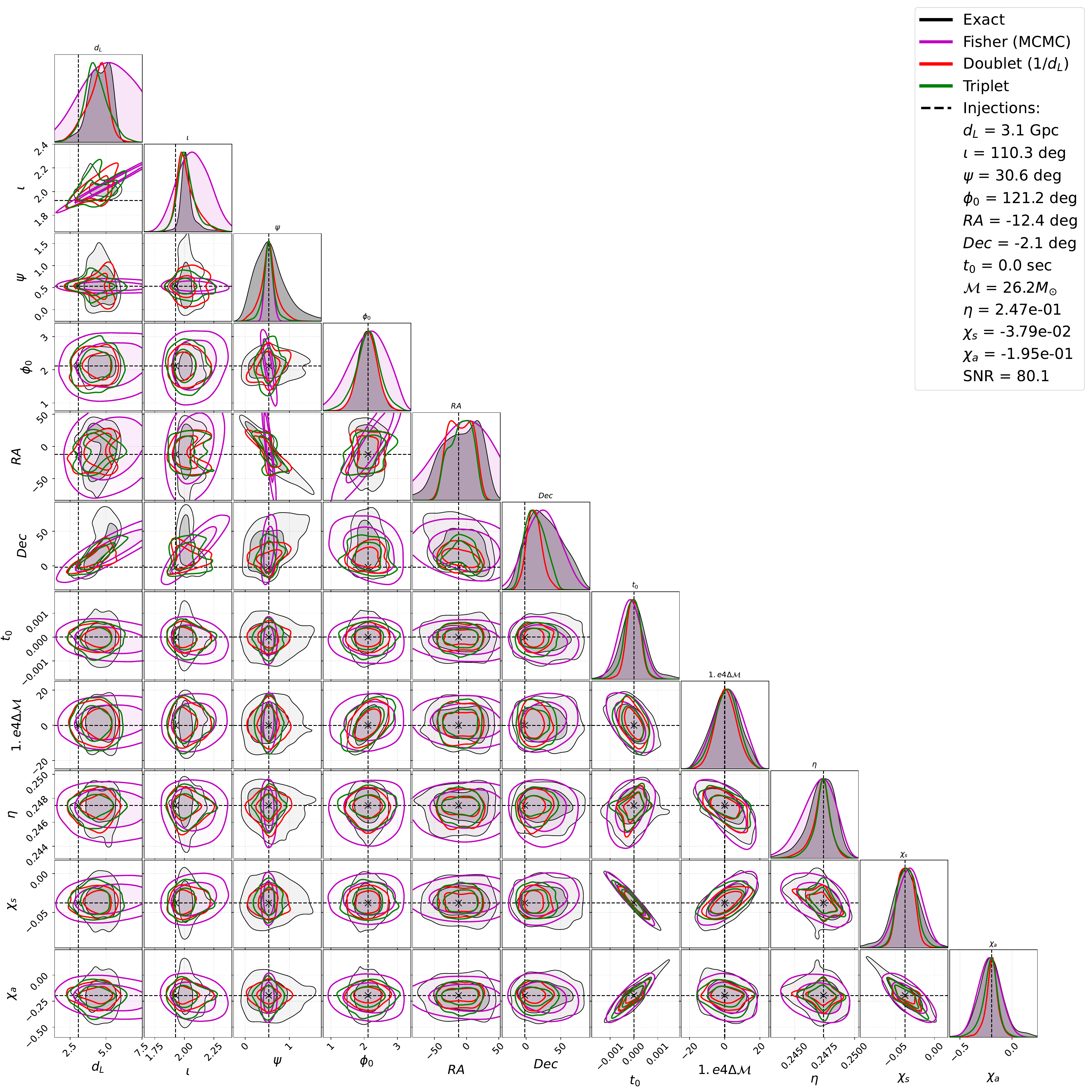}
    \caption{Same as Figure \ref{fig:corner-example-1} for a third, different source.}
    \label{fig:corner-example-3}
\end{figure}

In Figures~\ref{fig:corner-example-1}, \ref{fig:corner-example-2} and \ref{fig:corner-example-3} we illustrate three representative corner plots of three GW events, obtained with the different posterior approximations. These are sources with different SNRs: 14, 28 and 80. The other injection parameters are listed in the plot legends. Note that in some cases with strong multimodality in some parameters, such as $\psi$, RA and Dec, the different approximations often capture only the main peak, leading to an underestimation of the parameter uncertainties in those cases. Due to the high precision in the measurements of the chirp mass ${\cal M}$, we define the useful quantity $10^4 \Delta {\cal M} \equiv 10^4 ({\cal M} - {\cal M}_{\rm injected})$. We also note that the smoothing of the MCMC contours sometimes result in the plots slightly crossing the hard prior boundaries.

\section{Waveforms used in GWDALI 1.0}\label{app:waveforms}

\begin{figure}[t]
    \centering
    \includegraphics[width=.75\linewidth]{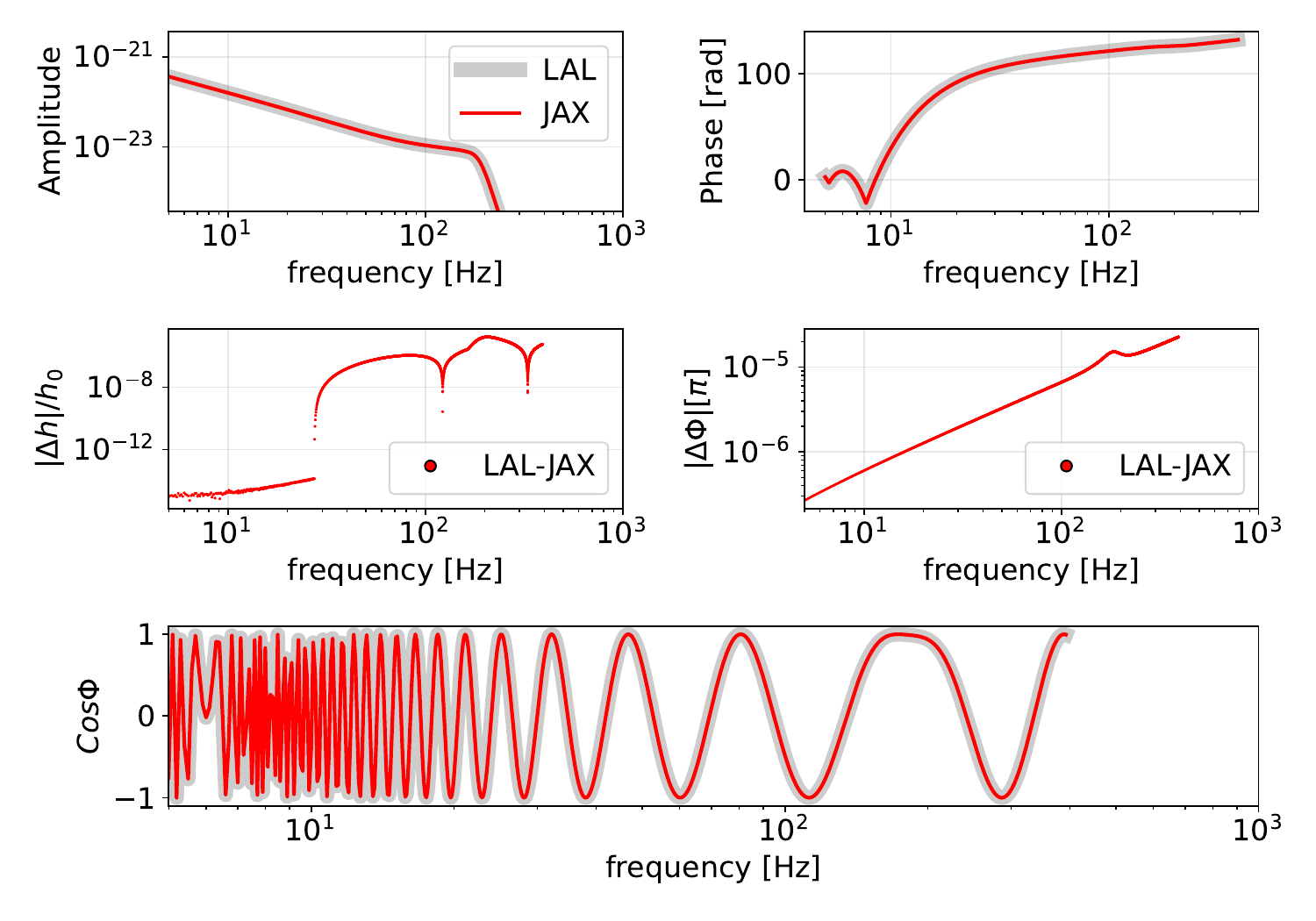}
    \caption{Comparison of the waveform IMRPhenomD obtained from \texttt{LAL} and \texttt{GWDALI} implemented via \texttt{JAX.numpy}.}\label{fig:PhenomD}
\end{figure}

\begin{figure}
    \centering
    \includegraphics[width=.75\linewidth]{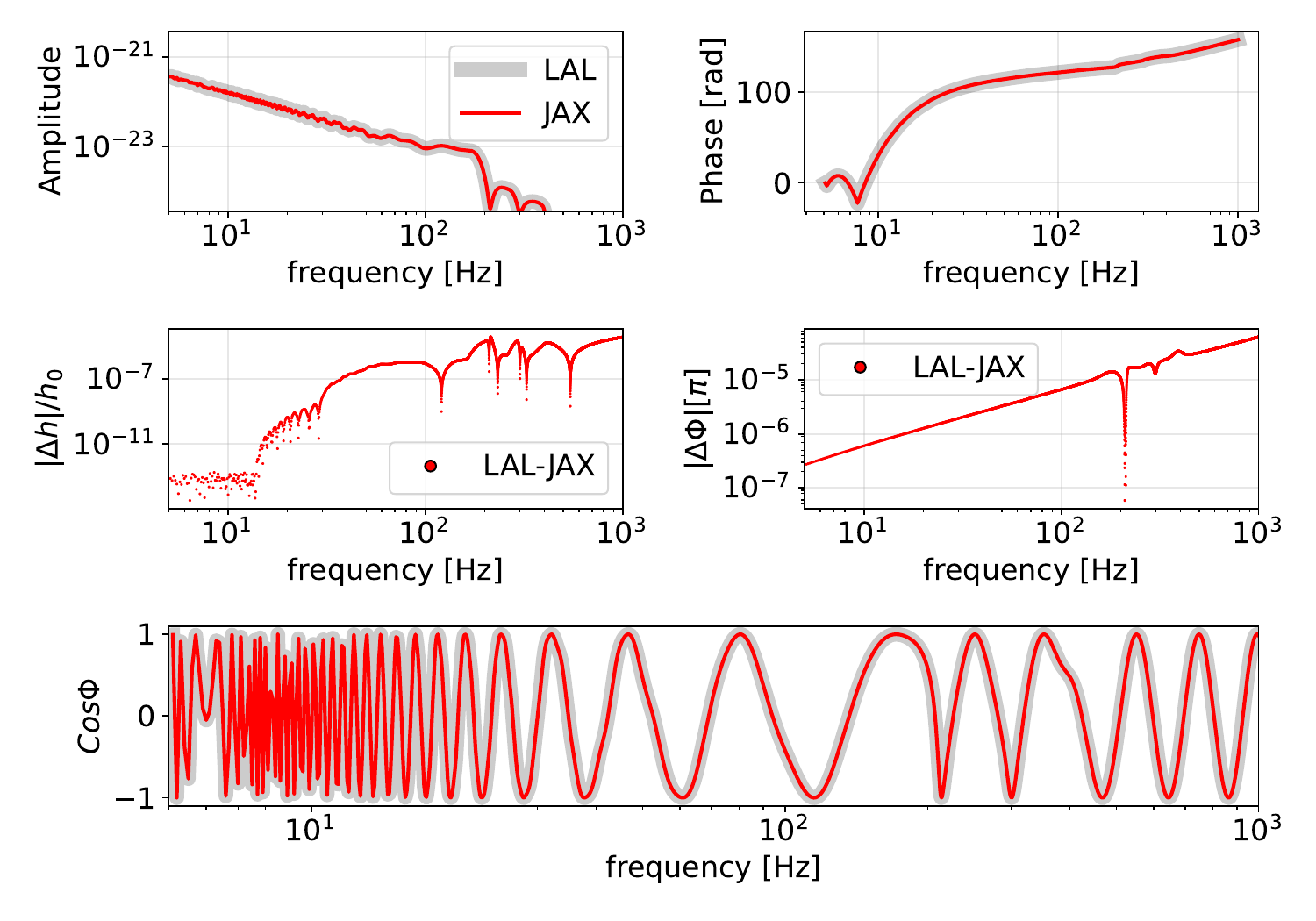}
    \caption{Comparison of the waveform IMRPhenomHM obtained from \texttt{LAL} and \texttt{GWDALI} implemented via \texttt{JAX.numpy}.}\label{fig:PhenomHM}
\end{figure}

In order for \texttt{GWDALI 1.0} to fully support automatic-differentiation, it was necessary to rewrite the gravitational-waveform models that had been employed in the previous version of the code. These models were re-implemented in a clean and transparent way using python–\texttt{JAX}, ensuring compatibility with \texttt{JAX}’s differentiation engine.
To validate our implementation, we compared the newly developed waveforms against the corresponding models provided by the widely used \texttt{LAL} libraries. In particular, we present in the Figures \ref{fig:PhenomD} and \ref{fig:PhenomHM}  the results for two waveform approximants: \texttt{IMRPhenomD} and \texttt{IMRPhenomHM}. For each case, we analyze the deviations in both the amplitude and the phase with respect to the \texttt{LAL} versions, demonstrating the level of agreement between our \texttt{JAX}-based implementation and the standard reference (see \cite{Iacovelli:2022mbg} for similar waveform implementations using \texttt{JAX} on \texttt{GWFAST-WF4Py}).

\begin{figure}[t]
    \centering
    \includegraphics[width=.8\linewidth]{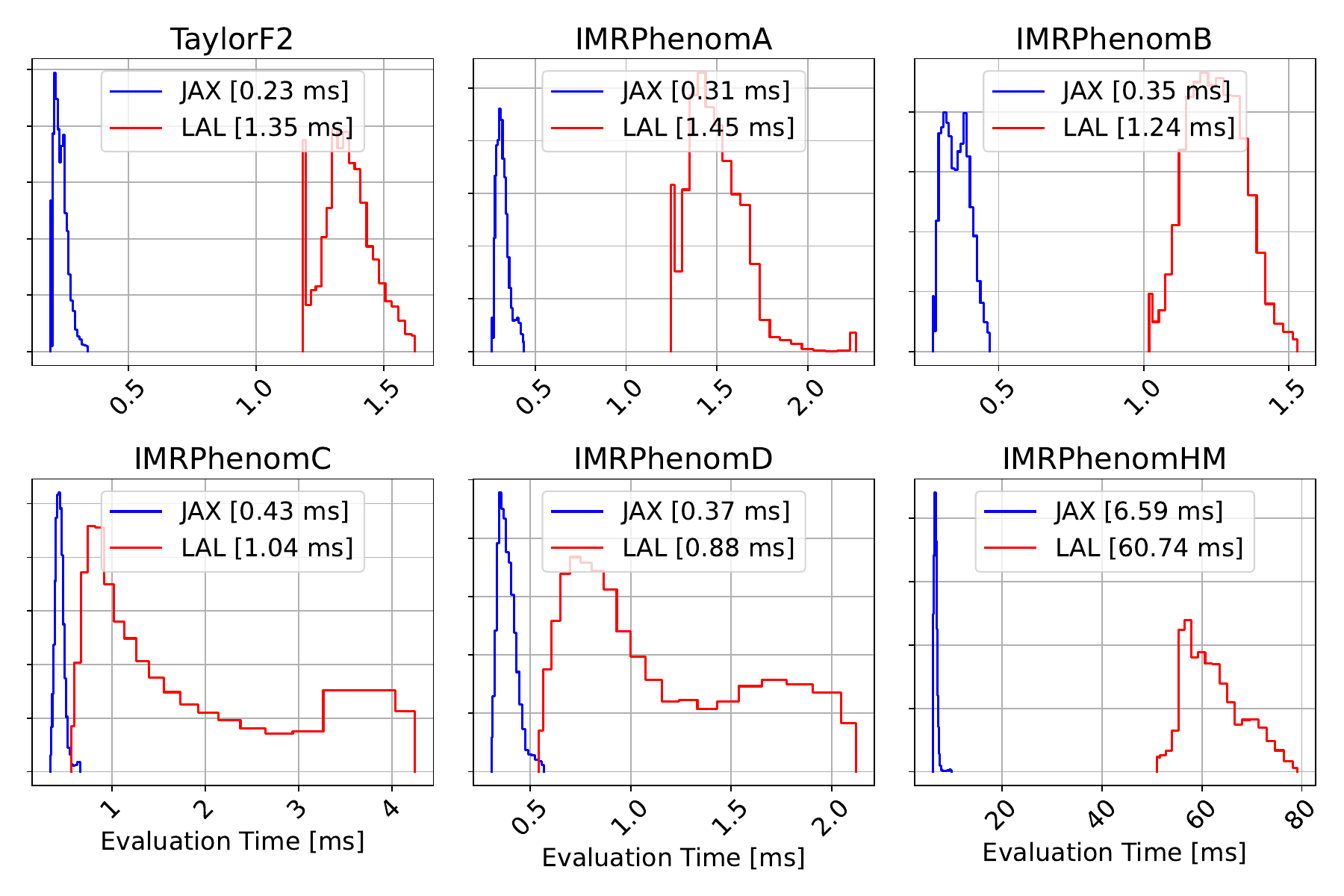}
    \caption{Elapsed time to call a single waveform in miliseconds using \texttt{JAX} (blue) and \texttt{LAL} (red). We also show in the legends the elapsed mean time to compute the waveforms. Histograms was build from 10,000 random GW sources.}\label{fig:WfTime}
\end{figure}

In addition to the concordance plots that compare our waveform implementations with those provided by \texttt{LAL}, Figure~\ref{fig:WfTime} shows the distribution of waveform evaluation times in \texttt{GWDALI~1.0}. This distribution was obtained from 10,000 calls, each corresponding to a different GW source, using all our waveform models implemented in python–\texttt{JAX}. As discussed previously, our implementations achieve faster evaluation times than the python-C interface of \texttt{LAL} counterparts, mainly due to the just-in-time compilation provided by \texttt{JAX}. In particular, for the \texttt{IMRPhenomHM}, it is almost 10 times faster than \texttt{LAL}.

\section{DALI performance for $\eta$ and $\delta_M$}\label{app:eta_vs_dM}

\begin{figure}[t]
    \centering
    \includegraphics[width=.6\linewidth]{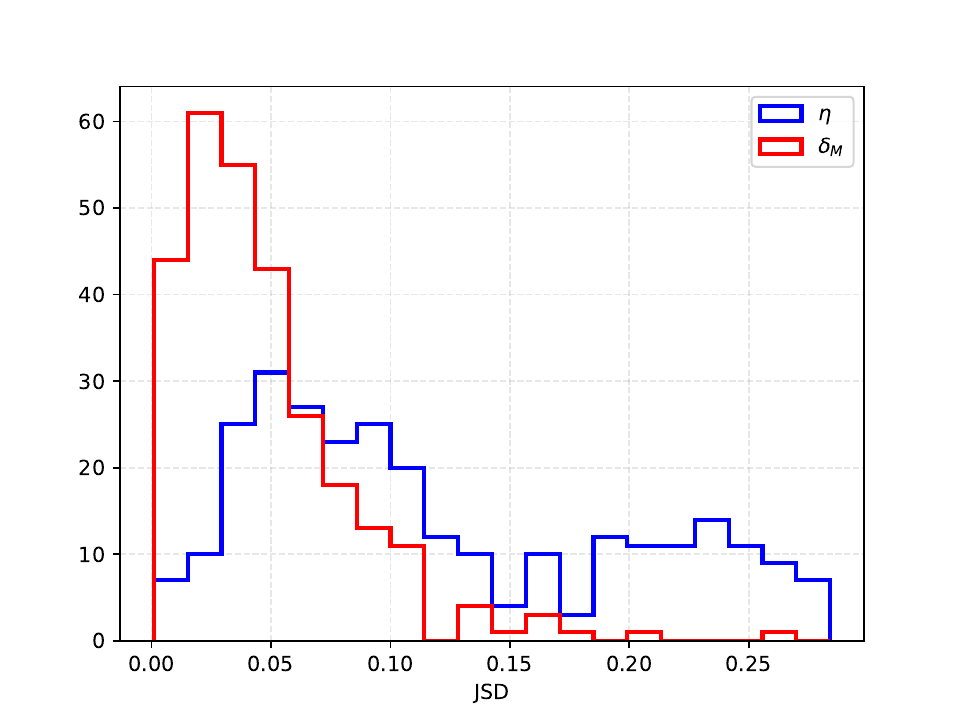}
    \caption{Distribution, for our 300 injections, of the JSD between the doublet-DALI and Exact posteriors with either $\eta$ or $\delta_M$ as the second free parameter for the masses. Using $\delta_M$ leads to better accuracy in general.} \label{fig:JSD_eta_vs_dM}
\end{figure}

Here we illustrate the DALI performance differences between using the symmetric mass ratio $\eta$ or the asymmetric mass parameter $\delta_M$. Figure~\ref{fig:JSD_eta_vs_dM} shows that using $\delta_M$ as the free parameter (and converting the samples \emph{a posteriori} to $\eta$ if so desired) leads on average to smaller JSD values. This Figure should be compared with the bottom left panel of Figure~\ref{fig:JSD-1D}.
    
\section{Tensors Transformations}\label{app:tns_transf}

When dealing with higher-order structures such as the Doublets, it is often advantageous to work with the inverse luminosity distance, $1/d_L$, rather than directly with $d_L$.  This choice simplifies the behavior of the derivatives of the waveform and leads to more stable numerical evaluations.  

Nevertheless, the formalism does not depend on this specific parametrization.  If one wishes to work with any alternative function of the luminosity distance (for example $d_L^n$), it is sufficient to compute the tensors in one convenient variable and then map them consistently to the new variable using the transformation rules derived below.  


Let us denote
\begin{equation}
    X \equiv d_{L}, 
\qquad Y \equiv d_{L}^{n}, 
\qquad X' \equiv \frac{\partial d_{L}}{\partial Y},
\end{equation}
so that derivatives of the waveform with respect to $Y$ can be systematically expressed in terms of derivatives with respect to $X$:
\begin{align}
\partial_{Y}h &= X' \, \partial_{X}h, \\
\partial_{Y}^{2}h &= X'' \, \partial_{X}h + (X')^{2}\, \partial_{X}^{2}h, \\
\partial_{Y}^{3}h &= X''' \, \partial_{X}h 
+ 3(X'X'')\, \partial_{X}^{2}h 
+ (X')^{3}\, \partial_{X}^{3}h .
\end{align}


For compactness, we label the scalar products of derivatives of the waveform as
\begin{equation}
    (n|m) \;\;\longleftrightarrow\;\; 
    \big\langle \partial_{X}^{\,n} h \,\big|\, \partial_{X}^{\,m} h \big\rangle ,
\end{equation}
where $n$ and $m$ represent the total number of derivatives with respect to $Y$ acting on the left and right sides of the inner product, respectively.  A first derivative counts as $1$, a second derivative counts as $2$, and a third derivative counts as $3$.  For example, the symbol $(2|3)$ indicates a scalar product where the left side involves two derivatives with respect to $d_L$ and the right side involves three.  

Since the inner product is symmetric, $\langle A | B \rangle = \langle B | A \rangle$, we adopt the convention of writing the larger number of derivatives on the right.  
This avoids redundancy and makes the transformation rules easier to compare across different tensors.

\subsection*{Fisher Matrix}

The Fisher information matrix is defined as
\begin{equation}
    {F}_{i,j} \equiv \langle \partial_{i}h \,|\, \partial_{j}h \rangle .    
\end{equation}

Transformation rules are as follows:
\begin{align}
    (0|1) \quad & {F}_{i,Y} = X' \, {F}_{i,X}, \\
    (1|1) \quad & {F}_{Y,Y} = (X')^{2} \, {F}_{X,X}. \nonumber
\end{align}

\subsection*{Doublet}

\paragraph{Type $(1,2)$:}
\begin{equation}
    {D}^{(1,2)}_{i,jk} \equiv 
    \langle \partial_{i}h \,|\, \partial_{j}\partial_{k}h \rangle .
\end{equation}

Transformation rules:
\begin{align}
(0|0) \quad & \langle i \,|\, jk \rangle 
  \equiv \langle \partial_{i}h \,|\, \partial_{j}\partial_{k}h \rangle , \nonumber \\
(0|1) \quad & \langle i \,|\, jY \rangle 
  = X' \, \langle i \,|\, jX \rangle , \nonumber \\
(0|2) \quad & \langle i \,|\, YY \rangle 
  = X'' \, \langle i \,|\, X \rangle 
  + (X')^{2} \, \langle i \,|\, XX \rangle , \\
(1|0) \quad & \langle Y \,|\, jk \rangle 
  = X' \, \langle X \,|\, jk \rangle , \nonumber \\
(1|1) \quad & \langle Y \,|\, jY \rangle 
  = (X')^{2} \, \langle X \,|\, jX \rangle , \nonumber \\
(1|2) \quad & \langle Y \,|\, YY \rangle 
  = (X'X'') \, \langle X \,|\, X \rangle 
  + (X')^{3} \, \langle X \,|\, XX \rangle . \nonumber
\end{align}

\paragraph{Type $(2,2)$:}
\begin{equation}
    {D}^{(2,2)}_{ij,kl} \equiv 
\langle \partial_{i}\partial_{j}h \,|\, \partial_{k}\partial_{l}h \rangle .
\end{equation}

Transformation rules:
\begin{eqnarray}
    (0|0) & \quad\langle ij|kl\rangle & \equiv\langle\partial_{i}\partial_{j}h|\partial_{k}\partial_{l}h\rangle \nonumber \\
    (0|1) & \quad\langle ij|k\mathbf{Y}\rangle & =X'\langle ij|k\mathbf{X}\rangle \nonumber \\
    (0|2) & \quad\langle ij|\mathbf{YY}\rangle & =X''\langle ij|\mathbf{X}\rangle+(X')^{2}\langle ij|\mathbf{XX}\rangle \nonumber \\
    (1|0) & \quad\langle i\mathbf{Y}|kl\rangle & =X'\langle i\mathbf{X}|kl\rangle\\
    (1|1) & \quad\langle i\mathbf{Y}|k\mathbf{Y}\rangle & =(X')^{2}\langle i\mathbf{X}|k\mathbf{X}\rangle \nonumber \\
    (1|2) & \quad\langle i\mathbf{Y}|\mathbf{YY}\rangle & =X'\cdot X''\langle i\mathbf{X}|\mathbf{X}\rangle+(X')^{3}\langle i\mathbf{X}|\mathbf{XX}\rangle \nonumber \\
    (2|0) & \quad\langle\mathbf{YY}|kl\rangle & =X''\langle\mathbf{X}|kl\rangle+(X')^{2}\langle\mathbf{XX}|kl\rangle \nonumber \\
    (2|1) & \quad\langle\mathbf{YY}|k\mathbf{Y}\rangle & =X'\cdot X''\langle\mathbf{X}|k\mathbf{X}\rangle+(X')^{3}\langle\mathbf{XX}|k\mathbf{X}\rangle \nonumber \\
    (2|2) & \quad\langle\mathbf{YY}|\mathbf{YY}\rangle & =(X'')^{2}\langle\mathbf{X}|\mathbf{X}\rangle+2(X')^{2}\cdot X''\langle\mathbf{X}|\mathbf{XX}\rangle+(X')^{4}\langle\mathbf{XX}|\mathbf{XX}\rangle \nonumber
\end{eqnarray}

\subsection*{Triplet}

\paragraph{Type $(1,3)$:}

\begin{equation}
    {T}_{i,klm}^{(1,3)} \equiv\langle\partial_{i}h|\partial_{k}\partial_{l}\partial_{m}h\rangle \,.
\end{equation}

Transformation rules:
\begin{eqnarray}
    (0|0) & \quad \langle i|jkl\rangle & \equiv\langle\partial_{i}h|\partial_{k}\partial_{l}\partial_{m}h\rangle \nonumber \\
    (0|1) & \quad \langle i|jk\mathbf{Y}\rangle & =X'\langle i|jk\mathbf{X}\rangle \nonumber \\
    (0|2) & \quad \langle i|j\mathbf{YY}\rangle & =X''\langle i|j\mathbf{X}\rangle+(X')^{2}\langle i|j\mathbf{XX}\rangle \nonumber \\
    (0|3) & \quad \langle i|\mathbf{YYY}\rangle & =X'''\langle i|\mathbf{X}\rangle+3X'\cdot X''\langle i|\mathbf{XX}\rangle+(X')^{3}\langle i|\mathbf{XXX}\rangle\\
    (1|0) & \quad \langle\mathbf{Y}|jkl\rangle & =X'\langle\mathbf{X}|jkl\rangle \nonumber \\
    (1|1) & \quad \langle\mathbf{Y}|jk\mathbf{Y}\rangle & =(X')^{2}\langle\mathbf{X}|jk\mathbf{X}\rangle \nonumber \\
    (1|2) & \quad \langle\mathbf{Y}|j\mathbf{YY}\rangle & =X'\cdot X''\langle\mathbf{X}|j\mathbf{X}\rangle+(X')^{3}\langle\mathbf{X}|j\mathbf{XX}\rangle \nonumber \\
    (1|3) & \quad \langle\mathbf{Y}|\mathbf{YYY}\rangle & =X'\cdot X'''\langle\mathbf{X}|\mathbf{X}\rangle+3(X')^{2}\cdot X''\langle\mathbf{X}|\mathbf{XX}\rangle+(X')^{4}\langle\mathbf{X}|\mathbf{XXX}\rangle \nonumber
\end{eqnarray}

\paragraph{Type $(2,3)$:}
\smallskip

\begin{equation}
    {T}_{ij,klm}^{(2,3)} \equiv\langle\partial_{i}\partial_{j}h|\partial_{k}\partial_{l}\partial_{m}h\rangle \,.
\end{equation}

Transformation rules:
{\small
\begin{eqnarray}
    (0|0) & \quad \langle ij|klm\rangle & \equiv\langle\partial_{i}\partial_{j}h|\partial_{k}\partial_{l}\partial_{m}h\rangle \nonumber \\
    (0|1) & \quad \langle ij|kl\mathbf{Y}\rangle & =X'\langle ij|kl\mathbf{X}\rangle \nonumber \\
    (0|2) & \quad \langle ij|k\mathbf{YY}\rangle & =X''\langle ij|k\mathbf{X}\rangle+(X')^{2}\langle ij|k\mathbf{XX}\rangle \nonumber \\
    (0|3) & \quad \langle ij|\mathbf{YYY}\rangle & =X'''\langle ij|\mathbf{X}\rangle+3X'\cdot X''\langle ij|\mathbf{XX}\rangle+(X')^{3}\langle ij|\mathbf{XXX}\rangle \nonumber \\
    (1|0) & \quad \langle i\mathbf{Y}|klm\rangle & =X'\langle i\mathbf{X}|klm\rangle \nonumber \\
    (1|1) & \quad \langle i\mathbf{Y}|kl\mathbf{Y}\rangle & =(X')^{2}\langle i\mathbf{X}|kl\mathbf{X}\rangle \\
    (1|2) & \quad \langle i\mathbf{Y}|k\mathbf{YY}\rangle & =X'\cdot X''\langle i\mathbf{X}|k\mathbf{X}\rangle+(X')^{3}\langle i\mathbf{X}|k\mathbf{XX}\rangle \nonumber \\
    (1|3) & \quad \langle i\mathbf{Y}|\mathbf{YYY}\rangle & =X'\cdot X'''\langle i\mathbf{X}|\mathbf{X}\rangle+3(X')^{2}\mathbf{X}''\langle i\mathbf{X}|\mathbf{XX}\rangle+(X')^{4}\langle i\mathbf{X}|\mathbf{XXX}\rangle \nonumber \\
    (2|0) & \quad \langle\mathbf{YY}|klm\rangle & =X''\langle\mathbf{X}|klm\rangle+(X')^{2}\langle\mathbf{XX}|klm\rangle \nonumber \\
    (2|1) & \quad \langle\mathbf{YY}|kl\mathbf{Y}\rangle & =X'\cdot X''\langle\mathbf{X}|kl\mathbf{X}\rangle+(X')^{3}\langle\mathbf{XX}|kl\mathbf{X}\rangle \nonumber \\
    (2|2) & \quad \langle\mathbf{YY}|k\mathbf{YY}\rangle & =(X'')^{2}\langle\mathbf{X}|k\mathbf{X}\rangle+(X')^{2}\cdot X''[\langle\mathbf{XX}|k\mathbf{X}\rangle+\langle\mathbf{X}|k\mathbf{XX}\rangle]+(X')^{4}\langle\mathbf{XX}|k\mathbf{XX}\rangle \nonumber \\
    (2|3) & \quad \langle\mathbf{YY}|\mathbf{YYY}\rangle & =X''\cdot X'''\langle\mathbf{X}|\mathbf{X}\rangle+[(X')^{2}X'''+3X'\cdot(X'')^{2}]\langle\mathbf{X}|\mathbf{XX}\rangle+3X''\cdot(X')^{3}\langle\mathbf{XX}|\mathbf{XX}\rangle \nonumber \\
    & & +(X')^{3}\cdot X''\langle\mathbf{X}|\mathbf{XXX}\rangle+(X')^{5}\langle\mathbf{XX}|\mathbf{XXX}\rangle \nonumber
\end{eqnarray}
}

\smallskip
\paragraph{Type $(3,3)$:}

\begin{equation}
    {T}_{ijk,lmn}^{(3,3)} \equiv\langle\partial_{i}\partial_{j}\partial_{k}h|\partial_{l}\partial_{m}\partial_{n}h\rangle \,.
\end{equation}

Transformation rules:
{\small
\begin{eqnarray}    
    (0|0) & \quad \langle ijk|lmn\rangle & \equiv\langle\partial_{i}\partial_{j}\partial_{k}h|\partial_{l}\partial_{m}\partial_{n}h\rangle \nonumber \\
    (0|1) & \quad \langle ijk|lm\mathbf{Y}\rangle & =X'\langle ijk|lm\mathbf{X}\rangle \nonumber \\
    (0|2) & \quad \langle ijk|l\mathbf{YY}\rangle & =X''\langle ijk|l\mathbf{X}\rangle+(X')^{2}\langle ijk|l\mathbf{XX}\rangle \nonumber \\
    (0|3) & \quad \langle ijk|\mathbf{YYY}\rangle & =X'''\langle ijk|\mathbf{X}\rangle+3X'\cdot X''\langle ijk|\mathbf{XX}\rangle+(X')^{3}\langle ijk|\mathbf{XXX}\rangle \nonumber \\
    (1|0) & \quad \langle ij\mathbf{Y}|lmn\rangle & =X'\langle ij\mathbf{X}|lmn\rangle \nonumber \\
    (1|1) & \quad \langle ij\mathbf{Y}|lm\mathbf{Y}\rangle & =(X')^{2}\langle ij\mathbf{X}|lm\mathbf{X}\rangle \nonumber \\
    (1|2) & \quad \langle ij\mathbf{Y}|l\mathbf{YY}\rangle & =X'\cdot X''\langle ij\mathbf{X}|l\mathbf{X}\rangle+(X')^{3}\langle ij\mathbf{X}|l\mathbf{XX}\rangle \nonumber \\
    (1|3) & \quad \langle ij\mathbf{Y}|\mathbf{YYY}\rangle & =X'\cdot X'''\langle ij\mathbf{X}|\mathbf{X}\rangle+3(X')^{2}\cdot X''\langle ij\mathbf{X}|\mathbf{XX}\rangle+(X')^{4}\langle ij\mathbf{X}|\mathbf{XXX}\rangle \\
    (2|0) & \quad \langle i\mathbf{YY}|lmn\rangle & =X''\langle i\mathbf{X}|lmn\rangle+(X')^{2}\langle i\mathbf{XX}|lmn\rangle \nonumber \\
    (2|1) & \quad \langle i\mathbf{YY}|lm\mathbf{Y}\rangle & =X'\cdot X''\langle i\mathbf{X}|lm\mathbf{X}\rangle+(X')^{3}\langle i\mathbf{XX}|lm\mathbf{X}\rangle \nonumber\\
    (2|2) & \quad \langle i\mathbf{YY}|l\mathbf{YY}\rangle & =(X'')^{2}\langle i\mathbf{X}|l\mathbf{X}\rangle+(X')^{2}\cdot X''[\langle i\mathbf{X}|l\mathbf{XX}\rangle+\langle i\mathbf{XX}|l\mathbf{X}\rangle]+(X')^{4}\langle i\mathbf{XX}|l\mathbf{XX}\rangle \nonumber \\
    (2|3) & \quad \langle i\mathbf{YY}|\mathbf{YYY}\rangle & =\left[X''\cdot X'''\langle i\mathbf{X}|\mathbf{X}\rangle+3X'\cdot(X'')^{2}\langle i\mathbf{X}|\mathbf{XX}\rangle\right]+(X')^{2}\cdot X'''\langle i\mathbf{XX}|\mathbf{X}\rangle \nonumber \\
     & & +X''\cdot(X')^{3}\left[3\langle i\mathbf{XX}|\mathbf{XX}\rangle+\langle i\mathbf{X}|\mathbf{XXX}\rangle\right]+(X')^{5}\langle i\mathbf{XX}|\mathbf{XXX}\rangle \nonumber \\
    (3|0) & \quad \langle\mathbf{YYY}|lmn\rangle & =X'''\langle\mathbf{X}|lmn\rangle+3X'\cdot X''\langle\mathbf{XX}|lmn\rangle+(X')^{3}\langle\mathbf{XXX}|lmn\rangle \nonumber \\
    (3|1) & \quad \langle\mathbf{YYY}|lm\mathbf{Y}\rangle & =X'\cdot X'''\langle\mathbf{X}|lm\mathbf{X}\rangle+3(X')^{2}\cdot X''\langle\mathbf{XX}|lm\mathbf{X}\rangle+(X')^{4}\langle\mathbf{XXX}|lm\mathbf{X}\rangle \nonumber \\
    (3|2) & \quad \langle\mathbf{YYY}|l\mathbf{YY}\rangle & =\left[X''\cdot X'''\langle\mathbf{X}|l\mathbf{X}\rangle + 3X'\cdot(X'')^{2}\langle\mathbf{XX}|l\mathbf{X}\rangle\right]  \nonumber \\
    &&+(X')^{2}\cdot X'''\langle\mathbf{X}|l\mathbf{XX}\rangle+3(X')^{3}X''\langle\mathbf{XX}|l\mathbf{XX}\rangle \nonumber \\
    & & +X''\cdot(X')^{3}\langle\mathbf{XXX}|l\mathbf{X}\rangle+(X')^{5}\langle\mathbf{XXX}|l\mathbf{XX}\rangle \nonumber \\
    (3|3) & \quad \langle\mathbf{YYY}|\mathbf{YYY}\rangle & =(X''')^{2}\langle\mathbf{X}|\mathbf{X}\rangle+\left[6X'\cdot X''\cdot X'''\langle\mathbf{X}|\mathbf{XX}\rangle+(3X'\cdot X'')^{2}\langle\mathbf{XX}|\mathbf{XX}\rangle\right] \nonumber \\
    & & +2(X')^{3}X'''\langle\mathbf{X}|\mathbf{XXX}\rangle+6(X')^{4}\cdot X''\langle\mathbf{XX}|\mathbf{XXX}\rangle+(X')^{6}\langle\mathbf{XXX}|\mathbf{XXX}\rangle \nonumber
\end{eqnarray}
}

\bibliographystyle{JHEP} 
\bibliography{gwdali}

\end{document}